\documentclass[twocolumn]{aastex62}
\graphicspath{{./}{figures/}}

\received{Mar 15, 2021}
\revised{January 28, 2022}
\accepted{\today}
\submitjournal{ApJ}

\shorttitle{ALMA High-resolution Multiband Analysis for TW~Hya}
\shortauthors{Tsukagoshi et al.}

\begin{document}

\title{ALMA High-resolution Multiband Analysis for the Protoplanetary Disk around TW~Hya}

\correspondingauthor{Takashi Tsukagoshi}
\email{takashi.tsukagoshi.astro@gmail.com}

\author[0000-0002-6034-2892]{Takashi Tsukagoshi}
\affil{Division of Science, National Astronomical Observatory of Japan, Osawa 2-21-1, Mitaka, Tokyo 181-8588, Japan}

\author{Hideko Nomura}
\affil{Division of Science, National Astronomical Observatory of Japan, Osawa 2-21-1, Mitaka, Tokyo 181-8588, Japan}

\author{Takayuki Muto}
\affil{Division of Liberal Arts, Kogakuin University, 1-24-2 Nishi-Shinjuku, Shinjuku-ku, Tokyo, 163-8677, Japan}

\author{Ryohei Kawabe}
\affil{Division of Science, National Astronomical Observatory of Japan, Osawa 2-21-1, Mitaka, Tokyo 181-8588, Japan}

\author{Kazuhiro D. Kanagawa}
\affil{College of Science, Ibaraki University, Bunkyo 2-1-1, Mito, Ibaraki, 310-8512, Japan}

\author{Satoshi Okuzumi}
\affil{Department of Earth and Planetary Sciences, Tokyo Institute of Technology, 2-12-1 Ookayama, Meguro, Tokyo, 152-8551, Japan }

\author{Shigeru Ida}
\affil{Earth-Life Science Institute, Tokyo Institute of Technology, 2-12-1 Ookayama, Meguro, Tokyo 152-8550, Japan}

\author{Catherine Walsh}
\affil{School of Physics and Astronomy, University of Leeds, Leeds, LS2 9JT, UK}

\author{Tom~J. Millar}
\affil{Astrophysics Research Centre, School of Mathematics and Physics, Queen's University Belfast, University Road, Belfast BT7 1NN, UK}

\author{Sanemichi~Z. Takahashi}
\affil{Division of Science, National Astronomical Observatory of Japan, Osawa 2-21-1, Mitaka, Tokyo 181-8588, Japan}

\author{Jun Hashimoto}
\affil{Astrobiology Center, 2-21-1 Osawa, Mitaka, Tokyo 181-8588, Japan}

\author{Taichi Uyama}
\affil{Infrared Processing and Analysis Center, California Institute of Technology, Pasadena, CA 91125, USA}
\affil{NASA Exoplanet Science Institute, California Institute of Technology, Pasadena, CA 91125, USA}

\author[0000-0002-6510-0681]{Motohide Tamura}
\affil{Astrobiology Center, 2-21-1 Osawa, Mitaka, Tokyo 181-8588, Japan}
\affil{Subaru Telescope, National Astronomical Observatory of Japan, Osawa 2-21-1, Mitaka, Tokyo 181-8588, Japan}
\affil{Department of Astronomy, School of Science, University of Tokyo, Bunkyo, Tokyo 113-0033, Japan}



\begin{abstract}
We present a high-resolution (2.5 au) multiband analysis of the protoplanetary disk around TW Hya using ALMA long baseline data at Bands 3, 4, 6, and 7. 
We aim to reconstruct a high-sensitivity millimeter continuum image and revisit the spectral index distribution. 
The imaging is performed by combining new ALMA data at Bands 4 and 6 with available archive data.
Two methods are employed to reconstruct the images; multi-frequency synthesis (MFS) and the fiducial image-oriented method, where each band is imaged separately and the frequency dependence is fitted pixel by pixel.
We find that the MFS imaging with the second order of Taylor expansion can reproduce the frequency dependence of the continuum emission between Bands 3 and 7 in a manner consistent with previous studies and is a reasonable method to reconstruct the spectral index map. 
The image-oriented method provides a spectral index map consistent with the MFS imaging, but with a two times lower resolution.
Mock observations of an intensity model were conducted to validate the images from the two methods. 
We find that the MFS imaging provides a high-resolution spectral index distribution with an uncertainty of $<10$~\%.
Using the submillimeter spectrum reproduced from our MFS images, we directly calculated the optical depth, power-law index of the dust opacity coefficient ($\beta$), and dust temperature. 
The derived parameters are consistent with previous works, and the enhancement of $\beta$ within the intensity gaps is also confirmed, supporting a deficit of millimeter-sized grains within the gaps.
\end{abstract}



\section{Introduction} \label{sec:intro}

Protoplanetary disks surrounding young stars are the birthplace of planets.
Forming planets are thought to interact with the parent protoplanetary disk and cause various substructures, such as an inner hole, gaps and rings, and large-scale asymmetries.
High-resolution observations with radio interferometers, such as the Atacama Large Millimeter/submillimeter Array (ALMA), have revealed that dust substructures within protoplanetary disks are common and are rich in variety \citep[e.g.,][]{bib:andrews2018}.
Recent high-resolution ALMA observations with deep integrations have revealed au-scale dust substructures that may be caused by a forming planet and a surrounding circumplanetary disk \citep{bib:tsukagoshi2019b,bib:isella2019}.
Further observational constraints are essential for confirming the physical origins of these substructures.

The first steps towards forming a planet involves the coagulation and growth of dust grains. 
Hence, revealing the evolution of dust grains in protoplanetary disks is a key for understanding the origin and diversity of planetary systems, and observational constraints on the dust size distribution is crucial.
Theoretical models of dust transport, fragmentation, and size evolution predict that the average size of grains varies with the disk radius \citep{bib:dullemond2005}.
The picture of dust filtration by a forming planet in a protoplanetary disk assumes that a planet-induced gap filters large dust grains at the outer edge of the gap, while the remaining small grains pass across the gap \citep{bib:zhu2012}.
It is also suggested that the maximum grain size should be smaller by a factor of 100 inside the condensation front of water ice, i.e., the H$_2$O snow line \citep{bib:banzatti2015}.
Because the H$_2$O snow line may be a boundary that determines the type of planet formed \citep[e.g., terrestrial planets, gas giants, or icy giants;][]{bib:hayashi1981}, it is important to reveal the position of the snow line in the protoplanetary disk to understand the planetary formation process.

Multi-frequency observations at (sub)millimeter wavelengths are an effective way to measure the dust size distribution and obtain a high-sensitivity intensity image by increasing the total bandwidth.
The dust size distribution can be inferred by measuring the spectral index $\alpha$ at (sub)millimeter frequencies.
When the dust continuum emission at (sub)millimeter frequencies is optically thin, the frequency dependence of the dust mass opacity coefficient $\kappa_\nu$ is evident in the $\alpha$ profile.
The dust mass opacity coefficient is often described as having a power-law form ($\kappa_\nu \propto \nu^\beta$), and in the Rayleigh-Jeans limit, $\beta$ is related to $\alpha$ as $\beta=\alpha-2$.
It is known that $\beta$ is affected by the dust size; $\beta\sim1.7$ for sub-micron-sized interstellar grains, while it changes to $\beta\sim1$ or less owing to grain growth in protoplanetary disks \citep[e.g.,][]{bib:miyake1993}.
Therefore, multi-frequency observations at optically thin (sub)millimeter wavelengths are essential to reveal the dust size distribution of the disk.

High-resolution multi-frequency observations with the Atacama Large Millimeter/submillimeter Array (ALMA) have been conducted on protoplanetary disks to resolve the radial dependence of the dust size distribution \citep[e.g,][]{bib:alma2015,bib:dent2019,bib:huang2020,bib:long2020}.
There are several ways to concatenate multi-frequency data for imaging the combined intensity and producing the spectral index maps.
The first is an image-oriented method.
This is the traditional method, in which the intensity map at each band is created separately and the frequency dependence is fitted pixel-by-pixel using a power-law function.
This method requires matching the beam sizes across all images before the fitting.
Another method to concatenate multi-frequency data is multi-scale multi-frequency synthesis (multi-scale MFS) introduced by \citet{bib:rau2011}, in which all observed visibilities are concatenated to simultaneously create combined intensity and spectral index maps.
The visibility-domain operation can provide higher-resolution maps than those of the image-oriented method.
\citet{bib:rau2011} demonstrated that multi-scale MFS works well for the imaging of a compact, flat-spectrum source at lower frequencies ($\sim$1~GHz), motivated by the application to synchrotron emission.
The authors pointed out that the UV coverage has a significant impact on reconstructing the spectral index map of spatially extended emission.
On the other hand, the thermal continuum emission within the dust of a protoplanetary disk has a spectral slope $\alpha$ of 2--4 depending on the optical depth and dust mass opacity coefficient.
In addition, recent high-resolution observations have revealed that the disks are often spatially extended.
Therefore, it is worth validating whether MFS works well for reconstructing the (sub)millimeter continuum emission of a protoplanetary disk and its frequency dependence.

TW~Hya is a 0.8 $M_\sun$ T Tauri star surrounded by a gas-rich protoplanetary disk at a distance of 59.5 pc \citep[e.g.,][]{bib:gaia2016}.
The disk is almost face-on with an inclination angle of 5--6$\degr$ \citep{bib:huang2018,bib:teague2019}; thus, it is one of the best targets for investigating the radial structure of a protoplanetary disk.
The disk has been well resolved at (sub)millimeter, near-infrared, and optical wavelengths.
Multiple gap structures in the near-infrared scattered light have been reported \citep{bib:akiyama2015,bib:boekel2017}.
ALMA has also resolved gaps at (sub)millimeter wavelengths, and an inner disk with a size of $\sim$1~au has also been identified \citep{bib:andrews2016,bib:tsukagoshi2016,bib:huang2018}.
The features detected thus far within the protoplanetary disk are almost axisymmetric except for a moving surface brightness asymmetry, probably due to a disk shadow \citep{bib:debes2017} and a spiral pattern found in the CO gas \citep{bib:teague2019}.
Another asymmetric structure of the disk is a localized compact ($\sim$1~au) excess emission at millimeter wavelengths near the edge of the dust disk identified in high-sensitivity ALMA observations \citep{bib:tsukagoshi2019b}.
The origin of the emission feature remains unclear, but it may be caused by a circumplanetary disk surrounding a Neptune-mass planet or dust grains accumulated within a small-scale gas vortex.
According to a recent theoretical study, the emission feature could also be a dust-losing young planet that has already been formed \citep{bib:nayakshin2020}.

The dust size distribution of the TW~Hya disk has been inferred using high-resolution multi-frequency observations with ALMA \citep{bib:tsukagoshi2016,bib:huang2018}.
The observations have revealed that the spectral index $\alpha$ decreases toward the disk center, and there is an enhancement near the gap at 25~au.
This enhancement may be attributed to the dust filtration effect, in which the gap is deficient in large grains \citep{bib:zhu2012}.
However, there is still uncertainty on the radial variation of the $\alpha$ profile.
The UV sampling of our previous observations in 2015 was particularly sparse at $<200$~k$\lambda$, and the integration time was as short as $\lesssim40$~min \citep{bib:tsukagoshi2016}. Hence, the poorly sampled UV coverage makes the image reconstruction difficult because the synthesized beam shows a complicated sidelobe pattern.
This caused difficulties in the image reconstruction by CLEAN, which highly depends on the imaging parameters, such as the weighting and the scale parameters of multiscale cleaning.
Additional uncertainty arises from adopting only two ALMA bands to derive the spectral index.
A combination of more than two bands can better constrain the spectral index by improving the frequency leverage.
Most recently, \citet{bib:macias2021} presented an analysis of the spectral index distribution of TW~Hya's disk using sets of high-resolution ALMA data from Bands~3 to 7, and the variation of the spectral index within the analyzed frequency range was reported.
As they focused on the spectral indices between two adjacent bands, a high-sensitivity continuum image integrated over all the bands was not presented.

In this study, we attempt to reconstruct a higher-sensitivity millimeter continuum image and revisit the spectral index distribution of the TW~Hya disk using multiple sets of ALMA data at Bands~3, 4, 6, and 7.
Two imaging methods, MFS and the image-oriented method, are adopted to combine all the data and to derive the spectral index map.
The details of the observations and data reduction are presented in \S~\ref{sec:obs}.
In \S~\ref{sec:results}, the images of the combined intensity and the spectral index are shown, and we compare them from the viewpoint of the different imaging methods.
To validate our reconstructed images, we tested the imaging methods using simulations with a disk model in \S~\ref{sec:model}.
In \S~\ref{sec:discussion}, we compare our results with recent high-resolution spectral index profiles presented by \citet{bib:macias2021}. 
We also discuss the dust size distribution in the disk by deriving the distribution of the optical depth $\tau$, power-law index of the dust mass opacity coefficient $\beta$, and dust temperature $T_\mathrm{d}$.
Lastly, we present a summary of this paper in \S\ref{sec:summary}.

\section{Observations and Data Reduction} \label{sec:obs}
\begin{figure*}[htb]
\begin{center}
\plotone{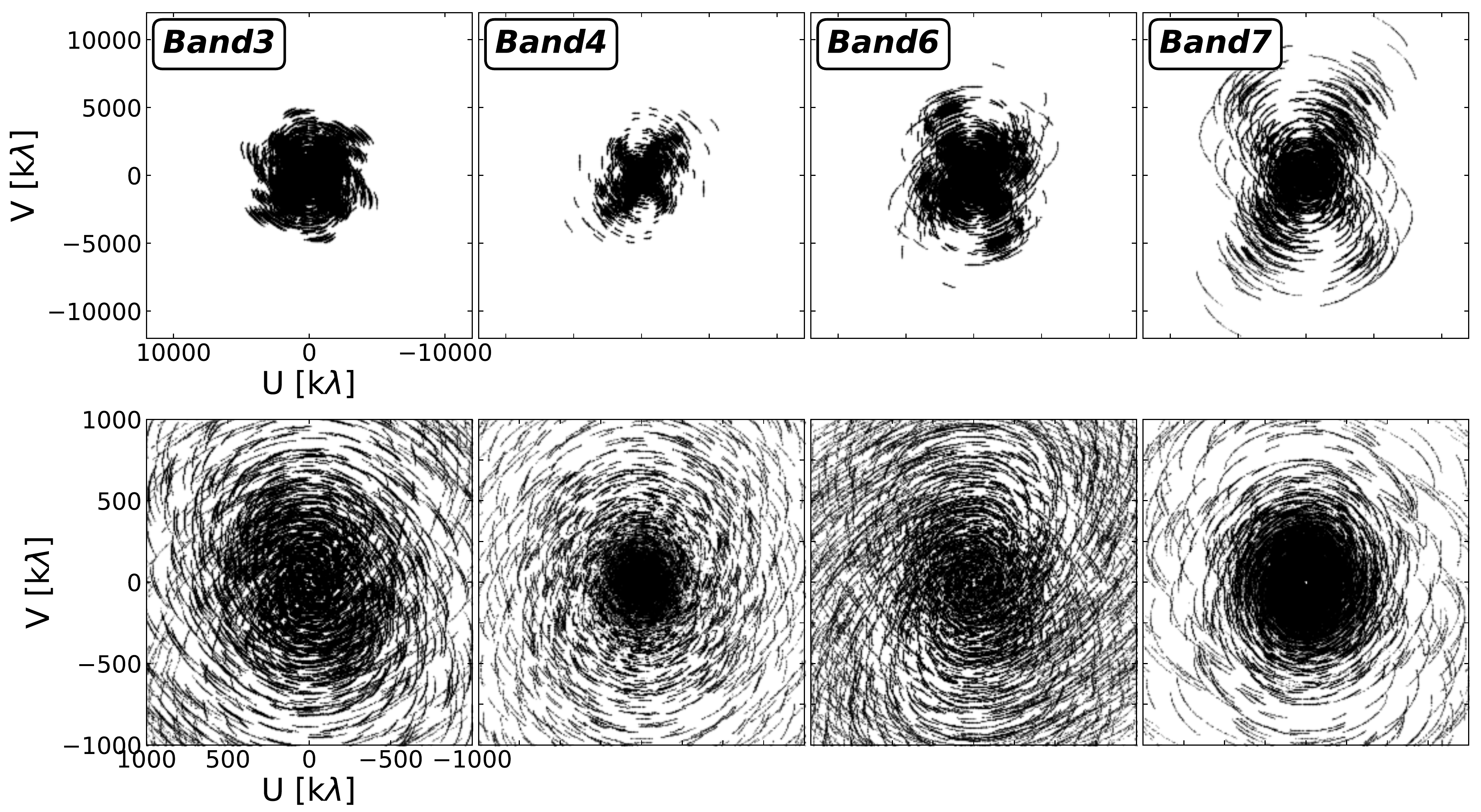}
\caption{Whole view of the UV coverage of the combined data of Bands 3, 4, 6 and 7 from left to right (top). Close-up view of the UV coverage inside $\pm$1000 k$\lambda$ (bottom).}\label{fig:uv}
\end{center}
\end{figure*}

In this study, we used sets of ALMA archive data at Bands~3, 4, 6, and 7 to reconstruct a high-sensitivity combined intensity map and a spectral index map covering these frequencies.
Here, we describe the details of our observations and data reduction.
The Band 4 and 6 data include our new observations, and the details of the observations are described in the following subsections.
All the ALMA measurement sets were reduced and calibrated using the Common Astronomical Software Application (CASA) package \citep{bib:mcmullin2007}.
The data IDs used in this study and their detailed information are listed in Table \ref{tab:ALMAdata}.
Figure\ \ref{fig:uv} shows the achieved UV coverage of each band's combined data.
The method for obtaining the combined intensity and spectral index maps is also described in the following subsection.

\begin{deluxetable*}{lcccccccc}\label{tab:ALMAdata}
\tabletypesize{\footnotesize}
\tablecaption{ALMA data employed in this study}
\tablehead{
\colhead{ID} &
\colhead{PI} &
\colhead{Date} &
\colhead{Configuration} &
\colhead{$L_\mathrm{min}$} &
\colhead{$L_\mathrm{max}$} &
\colhead{$t_\mathrm{integ}$} &
\colhead{$B_\mathrm{total}$} &
\colhead{CASA ver.} 
\\
\colhead{} &
\colhead{} &
\colhead{} &
\colhead{} &
\colhead{[m]} &
\colhead{[m]} &
\colhead{[min.]} &
\colhead{[MHz]} &
\colhead{}
}
\startdata
\multicolumn{9}{c}{Band 3} \\
\hline
2016.1.00229.S & Bergin,~E. & Aug 1, 2017 & C40-7 & 17 & 149 & 41 & 2293 & 4.7.2 \\
2018.1.01218.S & Macias,~E. & Jun 24--Jul 8, 2019 & C43-9/10 & 149 & 16196 & 209 & 7500 & 5.4.0 \\
\hline
\multicolumn{9}{c}{Band 4} \\
\hline
2015.A.00005.S & Tsukagoshi,~T. & Dec 2, 2015 & C36-8/7 & 17 & 10803 & 43 & 7500 & 4.5.0 \\
2015.1.00845.S & Favre,~C. & Apr 29, 2016 & C36-2/3 & 15 & 640 & 80 & 3750 & 4.5.3 \\
2015.1.00845.S & Favre,~C. & Jun 1, 2016 & C40-4 & 15 & 713 & 76 & 1875 & 4.5.3 \\
2016.1.00842.S & Tsukagoshi,~T. & Sep 28, 2017 & C40-6 & 19 & 1808 & 37 & 7500 & 4.7.0 \\
2016.1.00842.S & Tsukagoshi,~T. & Oct 21, 2016 & C40-8/9 & 41 & 14851 & 23 & 7500 & 4.7.2 \\
2016.1.00440.S & Teague,~R. & Oct 22, 2016 & C40-6 & 19 & 1400 & 141 & 1875 & 4.7.0 \\
\hline
\multicolumn{9}{c}{Band 6} \\
\hline
2013.1.00387.S & Guilloteau,~S. & May 13, 2015 & C34-3 & 21 & 558 & 47 & 1875 & 4.2.2 \\
2013.1.00114.S & O\"berg,~K. & Jul 19, 2014 & C34-4/5 & 34 & 650 & 43 & 938 & 4.2.2 \\
2015.A.00005.S & Tsukagoshi,~T. & Dec 2, 2015 & C36-8/7 & 17 & 10803 & 40 & 7500 & 4.5.0 \\
2016.1.00842.S & Tsukagoshi,~T. & May 15, 2017 & C40-5 & 15 & 1121 & 11 & 7500 & 4.7.2 \\
2017.1.00520.S & Tsukagoshi,~T. & Nov 20, 2017 & C43-8 & 92 & 8548 & 118 & 7500 & 5.1.1 \\
\hline
\multicolumn{9}{c}{Band 7} \\
\hline 
2015.1.00686.S & Andrews,~S. & Nov 23, 2015 & C36-8/7 & 17 & 14238 & 132 & 6094 & 4.5.0\\
2015.1.00308.S & Bergin,~E. & Mar 8, 2016 & C36-3 & 15 & 460 & 69 & 3750 & 4.5.2 \\
2016.1.00229.S & Bergin,~E. & Nov 23, 2016 & C40-4 & 15 & 704 & 49 & 3281 & 4.7.0 \\
2016.1.00440.S & Teague,~R. & Nov 27, 2016 & C40-3 & 15 & 704 & 48 & 1172 & 4.7.2 \\
2016.1.00464.S & Walsh,~C. & Dec 3, 2016 & C40-4 & 15 & 704 & 342 & 1875 & 4.7.2 \\
2016.1.01495.S & Nomura,~H. & Dec 6, 2016 & C40-3 & 15 & 704 & 43 & 1172 & 4.7.0 \\
2016.1.00629.S & Cleeves,~I. & Dec 30, 2016 & C40-3 & 15 & 460 & 84 & 2578 & 4.7.0 \\
2016.1.00311.S & Cleeves,~I. & May 21, 2017 & C40-5 & 15 & 390 & 48 & 1875 & 4.7.2 \\
\enddata
\end{deluxetable*}

\subsection{Band~3 data}
We have used archival data from two ALMA projects which were recently published by \citet{bib:macias2021}.
The details of the archive data and the CASA version used for the pipeline analysis are listed in Table~\ref{tab:ALMAdata}.
The pipeline script provided by ALMA was used for the initial data flagging and the calibration of the bandpass characteristics and the complex gain.
To concatenate the data obtained at different epochs, we first created a dirty map of each data set to determine the representative position of the emission, i.e., the center of the disk emission.
The dirty map was reconstructed with Briggs weighting with a robust parameter of 0.5, and the position of the emission peak was measured by a 2D Gaussian fitting to the emission using CASA {\it imfit}.
Subsequently, the field center of each measurement set was corrected to be the disk center by CASA {\it fixvis}.
Then, all the measurement sets were concatenated by CASA {\it concat} with a direction shift tolerance to be a single field center for correcting the proper motion of the source.

The concatenated visibilities were imaged using the CASA {\it tclean} task.
The CLEAN map was reconstructed by adopting the Briggs weighting with a robust parameter of 0.5.
We also employed the multiscale CLEAN algorithm with scale parameters of [0, 50, 150]~mas.
After the initial CLEAN map was reconstructed, we applied phase self-calibration to the concatenated data.
We adopted solution intervals varying from 1200 to 60~sec for the shorter baseline data and from 6000 to 900~sec for the longer baseline data.
The self-calibration started from longer solution intervals than the target scan to remove the systematic phase offsets between the concatenated measurement sets. 
Then, the self-calibration was stopped at the shortest solution interval where the signal-to-noise ratio is enough to solve above 2$\sigma$, i.e., only a small amount of visibilities were flagged out.
After the phase self-calibration was done, one round of amplitude self-calibration was applied with a solution interval for each observation period.
However, the image sensitivity was less affected by self-calibration so that the noise level of the final CLEAN map was 4.3~$\mu$Jy~beam$^{-1}$.
The beam size was $53.0\times50.7$~mas, with a position angle of $-7.7\degr$.

\subsection{Band~4 data}
Our ALMA Band 4 observations (2016.1.00842.S) were conducted on 2016 September 28, 2016, with array configuration C40-9 and on October 21, 2016, with C40-6. 
The total integration times were 12~min and 38~min, respectively. 
In addition to the observed data set, we employed ALMA archive data (2015.A.00005.S, 2015.1.00845.S, and 2016.1.00440.S) and concatenated them to obtain better sensitivity and UV coverage.

The initial flagging and calibrations were performed by using the pipeline scripts, and the calibrated visibilities were concatenated using the same procedure as for the Band 3 data.
The CLEAN map of the combined measurement set was reconstructed by adopting Briggs weighting with a robust parameter of 0.
We employed a multiscale option with scale parameters of [0, 50, 150] mas.
The self-calibration in phase was applied with solution intervals of 3600, 900, 300, and 150~sec, and followed by the amplitude self-calibration with intervals of each observation period.
The spatial resolution of the final CLEAN map at Band~4 was 85.1$\times$50.4~mas with a position angle of 45.4$\degr$, and the noise level of the self-calibrated CLEAN map was 7.8~$\mu$Jy~beam$^{-1}$.

\subsection{Band~6 data}
Our Band~6 observations were carried out on May 15, 2017, with array configuration C40-5 (2016.1.00842.S) and in the period from 2017 November 20 to 25, 2017, with C43-8 (2017.1.00520.S) during ALMA cycles 3 and 4. 
A description of the observations and the obtained image has already been published \citep{bib:tsukagoshi2019b}.
To improve the sensitivity, we obtained archive data 2013.1.00114.S, 2013.1.00387.S, and 2015.A.00005.S and concatenated them with the observed data to create the final Band~6 image.

After the initial data flagging and calibrations using the pipeline script, the same procedure as for the Band 3 data was applied to concatenate the calibrated measurement sets.
The imaging procedure was the same as that for the Band~3 data, except for some imaging parameters.
The CLEAN map was reconstructed using Briggs weighting with a robust parameter of 0.5.
The scale parameters for the multiscale CLEAN were set to [0, 42, 126] mas.
The phase-only self-calibration was applied varying the solution interval from 3600 to 120~sec, and was followed by the amplitude self-calibration.
The noise level of the final CLEAN image was 8.1~$\mu$Jy~beam$^{-1}$.
The beam size of the final image was 46.9$\times$40.6 mas with a position angle of 86.4$\degr$.

\subsection{Band~7 data}
To create a high-resolution Band 7 image, we have used eight sets of ALMA archive data presented in \citet{bib:tsukagoshi2019b}, including the highest resolution data obtained by \citet{bib:andrews2016}.
The data reduction and imaging was performed with the same procedure as for the other bands, except for some imaging parameters.
We employed the Briggs weighting with a robust $-1.0$ for reconstructing the CLEAN map.
The phase-only self-calibration was applied varying the solution interval from 7200 to 1200~sec.
With the phase-only self-calibration, the image noise level was improved from 124 to 32.8~uJy~beam$^{-1}$, corresponding to an improvement in the SNR from $\sim$18 to $\sim$62.
Then, the amplitude self-calibration was performed with a solution interval of each observation period.
The noise level and the beam size of the final CLEAN image are 21.8~$\mu$Jy~beam$^{-1}$ and 36.4$\times$28.9~mas with a position angle of 69.9$\degr$, respectively. 
The details of the data reduction were also described in \citet{bib:tsukagoshi2019b}.

\subsection{Reconstruction of the intensity and spectral index maps from all bands data}
To combine the entire data set from Bands~3 to 7, we first corrected the proper motion by aligning the field center in the same manner for each band data.
The disk center was derived by 2D Gaussian fitting to the bright part of the emission in the CLEAN map of each band.
The field center of the measurement set at each band was updated to be the disk center.
Then, all the measurement sets were concatenated using {\it concat} with a direction tolerance being a single field.
With this concatenated measurement set, we reconstructed the intensity and spectral index maps at the central frequency using the two methods described in the following subsections.
To match the minimum and maximum UV length between all band data, we employed the data in the baseline range of 14--5100 k$\lambda$.
We used CASA version 6.2 for reconstructing the images.

\subsubsection{Image-oriented method}
Before making the maps, the concatenated measurement set was first divided into each band, and CLEAN images were made with Briggs weighting with a robust parameter of 0.
We employed the same image size and cellsize to directly apply the mathematical operation to the images.
The multiscale CLEAN algorithm was also employed with scales of 0, 54, 162~mas.
All the reconstructed CLEAN images were convolved to have a circular beam with a full width at half maximum (FWHM) of 108~mas ($\sim6.4$~au), which is the largest beam major axis among the CLEAN images.

For each pixel in the convolved images, we fit a power-law function $I=I_0 (\nu/\nu_0)^{I_\alpha}$ along the frequency axis.
Here, $I_0$ is the intensity at the central frequency $\nu_0=$221~GHz.
Note that image pixels where the emission is higher than 5$\sigma$ are used for the fit.
The noise levels of the CLEAN maps are 4.3, 9.4, 13, and 49~$\mu$Jy~beam$^{-1}$ for Bands~3, 4, 6, and 7, respectively.
Fitting was performed using {\it curve\_fit} in the {\it scipy} package\footnote{https://scipy.org} \citep{bib:scipy2020}.

According to the ALMA proposer's guide, the uncertainties of the absolute flux calibration of ALMA are 5, 5, 10, and 10\% for Bands~3, 4, 6, and 7, respectively.
This corresponds to the fitting error of the spectral index with the image-oriented method being less than 0.01.
Note that the uncertainty of the absolute flux calibration is lower than the above value because we combine some measurement sets for each band.

\subsubsection{Multi-scale and multi-frequency synthesis}
To create the combined intensity and spectral index maps, we also employed a multi-scale multi-frequency synthesis method (hereafter multi-scale MFS) implemented in the CASA {\it tclean} task \citep[deconvolver=mtmfs;][]{bib:rau2011}.
 
In this method, the images are reconstructed by simultaneously solving the CLEAN components in the spatial and spectral regimes.
In particular, the MFS method solves the frequency dependence of the intensity by adopting the Taylor expansion of the following equation,

\begin{eqnarray}
I_\nu && = I_{\nu_0} \left( \frac{\nu}{\nu_0} \right) ^{I_\alpha + I_\beta \log \left( \frac{\nu}{\nu_0} \right)} \label{eq:mfs1} \\
      && \sim I_0 + I_1 \left( \frac{\nu-\nu_0}{\nu_0} \right) + I_2 \left( \frac{\nu-\nu_0}{\nu_0} \right)^2 + \dots \ . 
\end{eqnarray}
Here, $I_{\nu_0}$ is the intensity value at the representative frequency $\nu_0$, and $I_\alpha$ and $I_\beta$ are the values of the power-law index and the curvature of the frequency dependence, respectively.
The Taylor coefficients $I_\mathrm{n}$ ($n=0,1,2,\dots$) were determined via the deconvolution process.
If we take the first order of the Taylor expansion, the first two coefficients $I_0$ and $I_1$ correspond to $I_0 = I_{\nu_0}$ and $ I_1=I_\alpha I_{\nu_0}$, and thus $I_{\nu_0}$ and $I_\alpha$ can be calculated from the coefficients.
For the second order of the Taylor expansion, $I_\beta$ can be obtained using the third coefficient 
\begin{equation}\label{eq:mfs2}
I_2 = \biggl( \frac{I_{\alpha} (I_{\alpha}-1)}{2} + I_\beta \biggr) I_{\nu_0}\ .
\end{equation}
The polynomial approximation of the power-law function is a source of errors.
Although increasing the number of Taylor terms would be better for reproducing the power-law dependence of the frequencies, the use of too many terms could increase the critical errors for noisy data because of the increasing number of free parameters.
In addition, the total frequency coverage of available images with respect to the representative frequency, i.e., the bandwidth ratio, could be a source of errors.
This is because the wider the bandwidth, the more Taylor terms are required to reproduce the power-law dependence.

The concatenated measurement set with data from all bands was imaged by adopting the {\it mtmfs} option in {\it tclean}, in which the number of Taylor coefficients is controlled by the {\it nterms} parameter; {\it nterms}=2 and {\it nterms}=3 mean that the frequency dependence is described by the Taylor expansion to the first and second orders, respectively.
We created maps with {\it nterms}=2,  3, and  4 because the frequency range is wide, with a value of 95--360~GHz.
The combined intensity map at a central frequency of 221~GHz and the spectral index map were reconstructed from all the calibrated visibilities using Briggs weighting with a robust parameter of 0.
The scale parameter of the multiscale CLEAN was set to [0, 54, 162]~mas.
The resolution of the final images was46.0$\times$42.5~mas with a position angle of 42.3$\degr$. 

Note that the uncertainty in the spectral index measurement with this method due to the absolute flux calibration is estimated to be less than 8\% from mock observations with an intensity model.
Moreover, the uncertainty does not affect the shape of the $I_\alpha$ profile, but the entire profile was scaled.
See \S~\ref{sec:model} for more detail.

\section{Results} \label{sec:results}
\begin{figure*}[htb]
\begin{center}
\epsscale{0.8}
\plotone{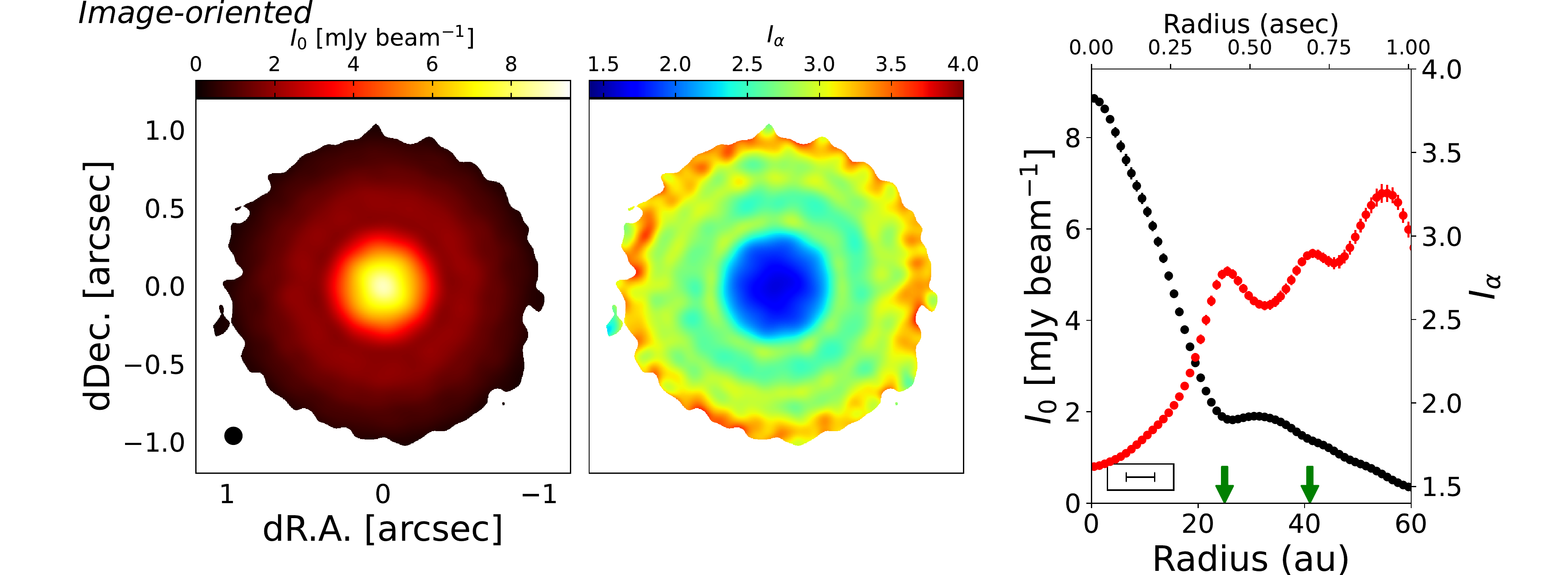}
\plotone{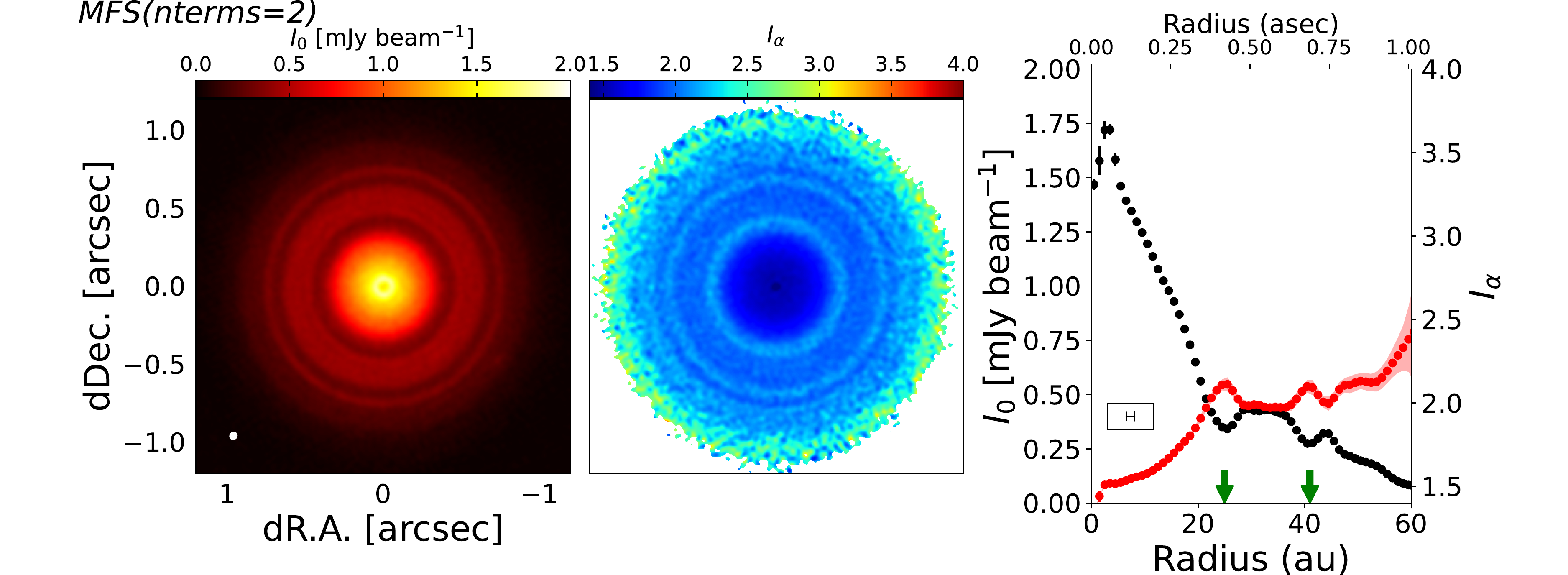}
\plotone{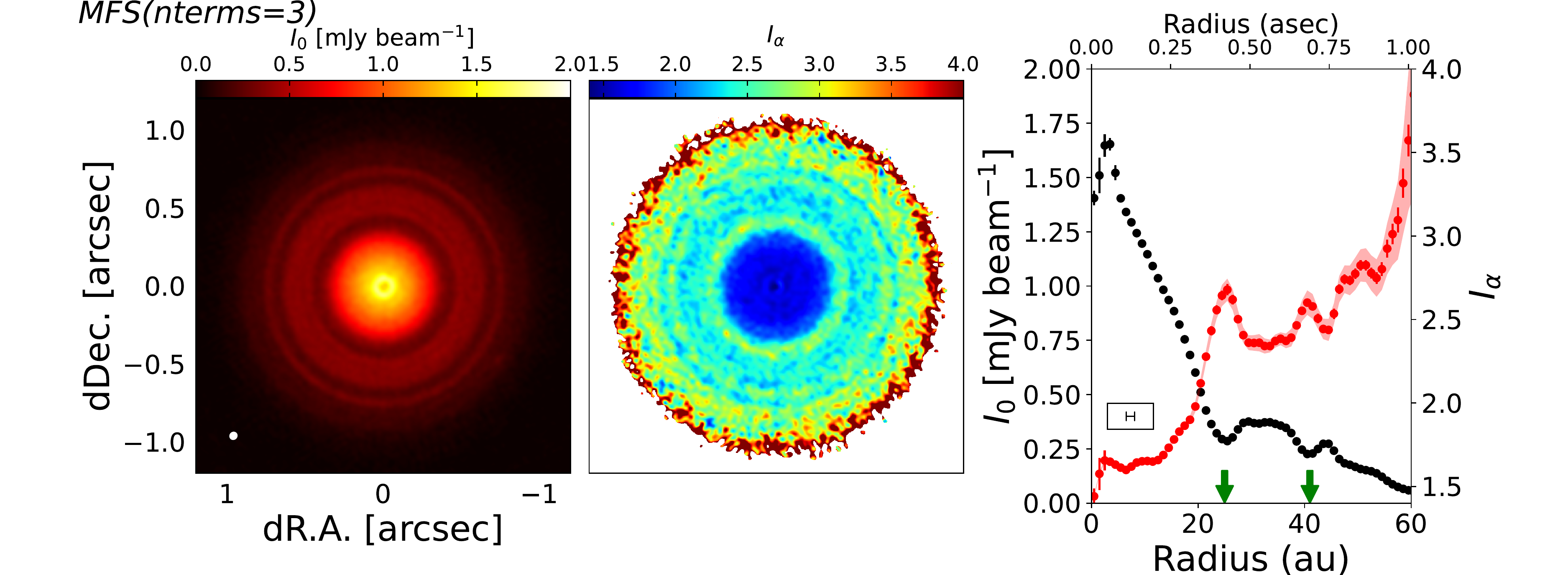}
\plotone{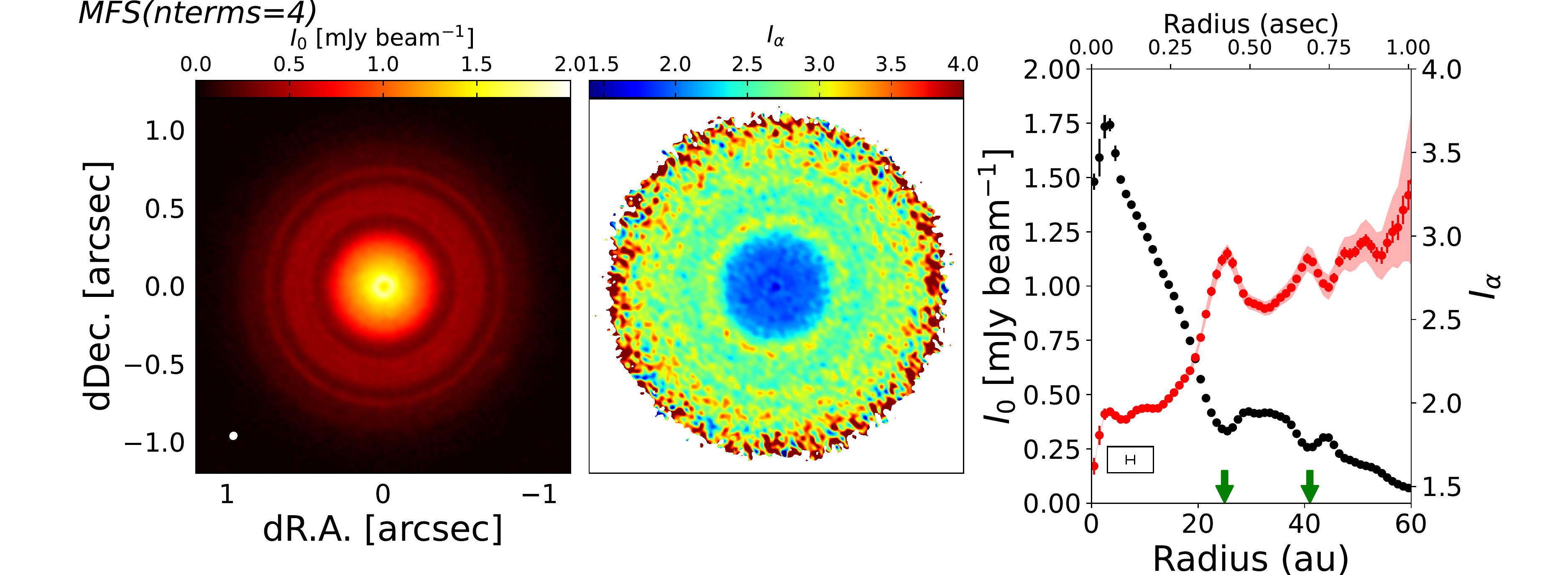}
\caption{Combined intensity (left) and spectral index $I_\alpha$ (center) maps. The circle at the bottom-left corner of the intensity panel shows the beam size. The righthand panel shows the radial profiles of the intensity (black) and $I_\alpha$ (red) azimuthally averaged after the image deprojection. The error bars show the standard error through the azimuthal averaging. The shaded region indicates the value of the $I_\alpha$ error map that CASA {\it tclean} provided. The bar at the bottom-left corner denotes the geometric mean of the beam size. The images reconstructed by the image-oriented analysis are shown in the top row, and those reconstructed by MFS are shown from second to the bottom. The noise of the intensity map of the image-oriented method is 95~$\mu$Jy~beam$^{-1}$, and that for the MFS maps of {\it nterms}=2--4 is 7.5~$\mu$Jy~beam$^{-1}$. The green arrows in the right panels indicate the positions of the 25 and 41~au gaps.}\label{fig:spindex}
\end{center}
\end{figure*}

Figure \ref{fig:spindex} (top) shows the intensity map at the central frequency (221~GHz) $I_0$ and the spectral index map $I_\alpha$ obtained using the image-oriented method. 
Although the beam size is almost doubled with respect to that of previous studies \citep{bib:andrews2016,bib:tsukagoshi2016,bib:huang2018}, the combined intensity map resolves the disk substructures, two clear gaps and an inner hole, as shown in the leftmost panel of Figure \ref{fig:spindex} (top).
The total flux density integrated over the disk emission is estimated to be 403~mJy.
The spectral index map shows the radial variation as previously reported \citep{bib:tsukagoshi2016,bib:huang2018}.
The spectral index is $\sim$3.0 near the outermost disk and decreases to less than 2 inside 20~au.
There are enhancements of the spectral index likely associated with the gaps in the intensity distribution at 25 and 42 au.
The rightmost panel of Figure\ \ref{fig:spindex}(top) shows the deprojected radial profiles of the intensity and the spectral index maps.
The error bars are determined by the standard error through the azimuthal averaging.
Note that for the deprojection, we employed an inclination of 7$\degr$ \citep{bib:qi2004}, which is 1--2$\degr$ larger than that determined by recent works \citep{bib:huang2018,bib:teague2019}.
This slight difference does not affect the profiles.

The $I_0$ and $I_\alpha$ maps reconstructed using MFS are also shown in Figure \ref{fig:spindex} (second to bottom).
The deprojected radial profiles are also shown in the right panels of the Figure.
The shaded region of the $I_\alpha$ profile shows the $I_\alpha$ error map, which is the outcome of the CASA {\it tclean} with the {\it mtmfs} option.
Evidently, with a higher spatial resolution of MFS than that of the image-oriented method, the $I_0$ map shows the disk substructures more clearly.
The intensity maps reconstructed with different {\it nterms} show no clear difference.
The image noise level of the maps is 7.5~$\mu$Jy~beam$^{-1}$.
The peak intensities are 1.79, 1.73, and 1.83~mJy~beam$^{-1}$, and the total flux densities are 526, 471, and 549~mJy for {\it nterms}=2, 3 and {\it 4}, respectively.
There is a slight difference in the measured flux densities, but it is less than 10\%.

The shape of the $I_\alpha$ profile reconstructed by MFS with {\it nterms}=2 is similar to that of the image-oriented method.
Starting from the outermost part, $I_\alpha$ gradually decreases toward $\sim$25~au with slight enhancements associated with the intensity gaps, suddenly drops near 20~au, and has a lower value in the innermost region.
However, the absolute value of $I_\alpha$ reconstructed with {\it nterms}=2 is much smaller than that of the image-oriented method.

In contrast to the {\it nterms}=2 case, the absolute value of $I_\alpha$ is similar to that for the {\it nterms}=3 and 4 cases.
Both cases show a similar $I_\alpha$ profile varying from $\sim$3 to $\lesssim$2 toward the disk center, whereas there is a slight difference between them ($\sim$0.3).
The $I_\alpha$ enhancements at the gaps are also observed to have a similar value to the image-oriented case.

\begin{figure}[htb]
\begin{center}
\plotone{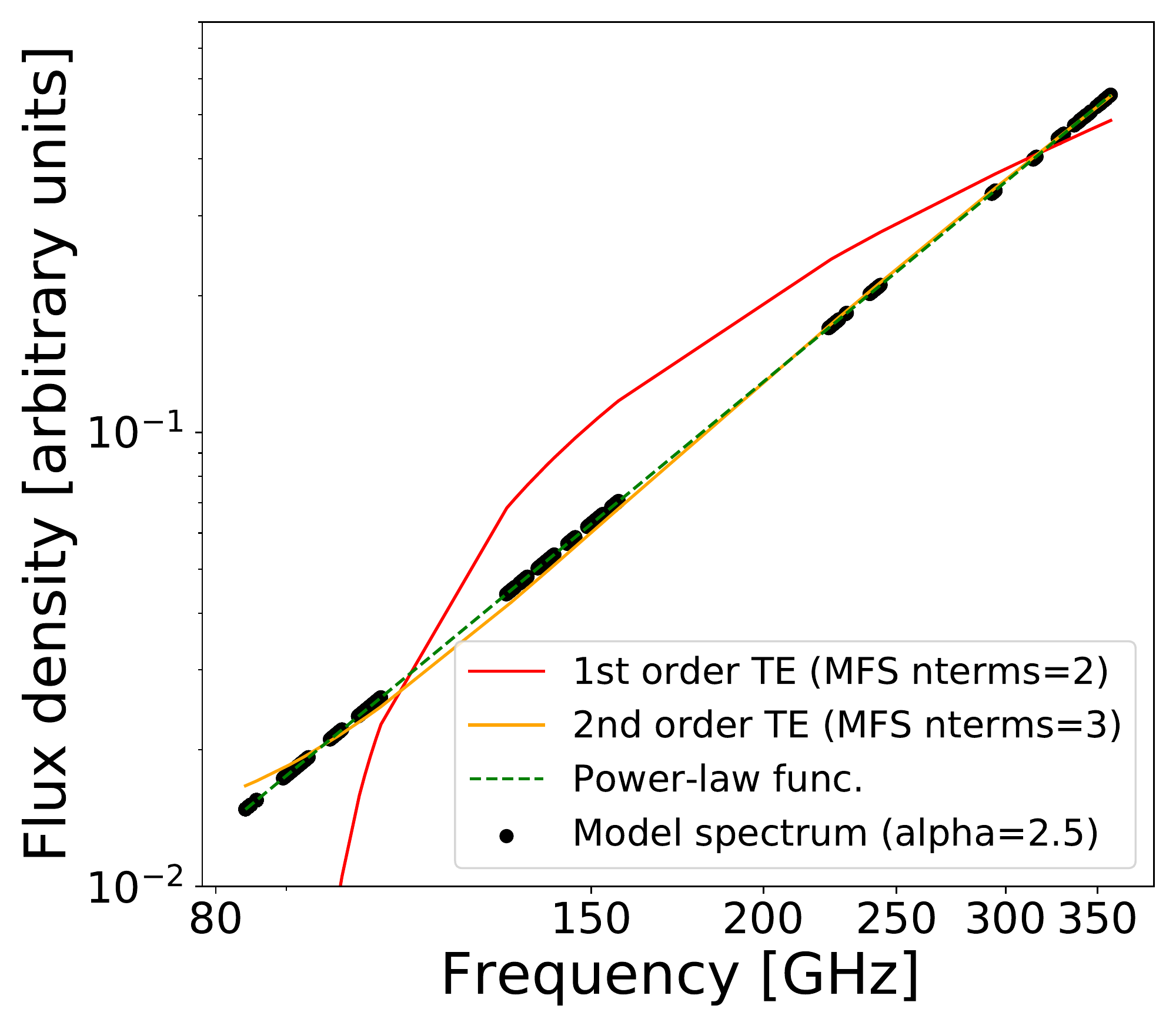}
\caption{Reproducibility of the submillimeter spectrum for different {\it nterms}. The flux densities are determined to follow a power-law index of 2.5, and the spectrum is sampled at the frequencies of the observation data. The solid lines show the fit using the equations for {\it nterms}=2 (red) and {\it nterms}=3 (yellow), while the fit using a power-law function is shown by the dotted line in green. The vertical axis shows the flux density in arbitrary units. `TE' is an abbreviation of Taylor expansion.}\label{fig:model_sed}
\end{center}
\end{figure}

To determine how the order of the Taylor expansion reproduces the power-law dependence across the observed frequencies, we performed a least-square fitting of the first and second orders of the Taylor expansion of a power-law function to a model spectrum with $I_\nu \propto \nu^{2.5}$ sampled at the observed frequencies.
A pure power-law function was also employed for the fitting as a reference.
The results of the fitting is shown in Figure~\ref{fig:model_sed}.
As mentioned previously, the first order of the Taylor expansion ({\it nterms}=2) is insufficient to reproduce the power-law form of the submillimeter spectrum between the observed bands.
In contrast, the second order of the Taylor expansion can almost reproduce the power-law dependence.
This indicates that at least the second order of the Taylor expansion is required to measure the spectral index between the observed bands with MFS.

\begin{figure*}[htb]
\begin{center}
\epsscale{1.0}
\plotone{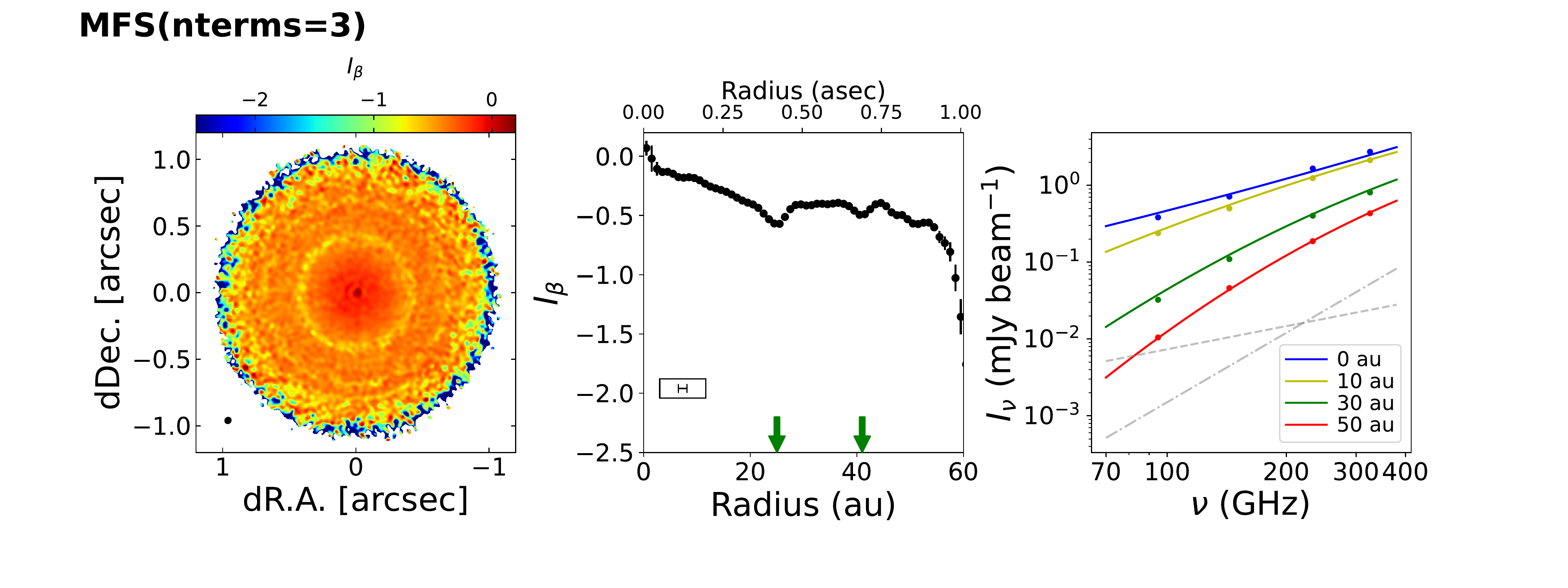}
\plotone{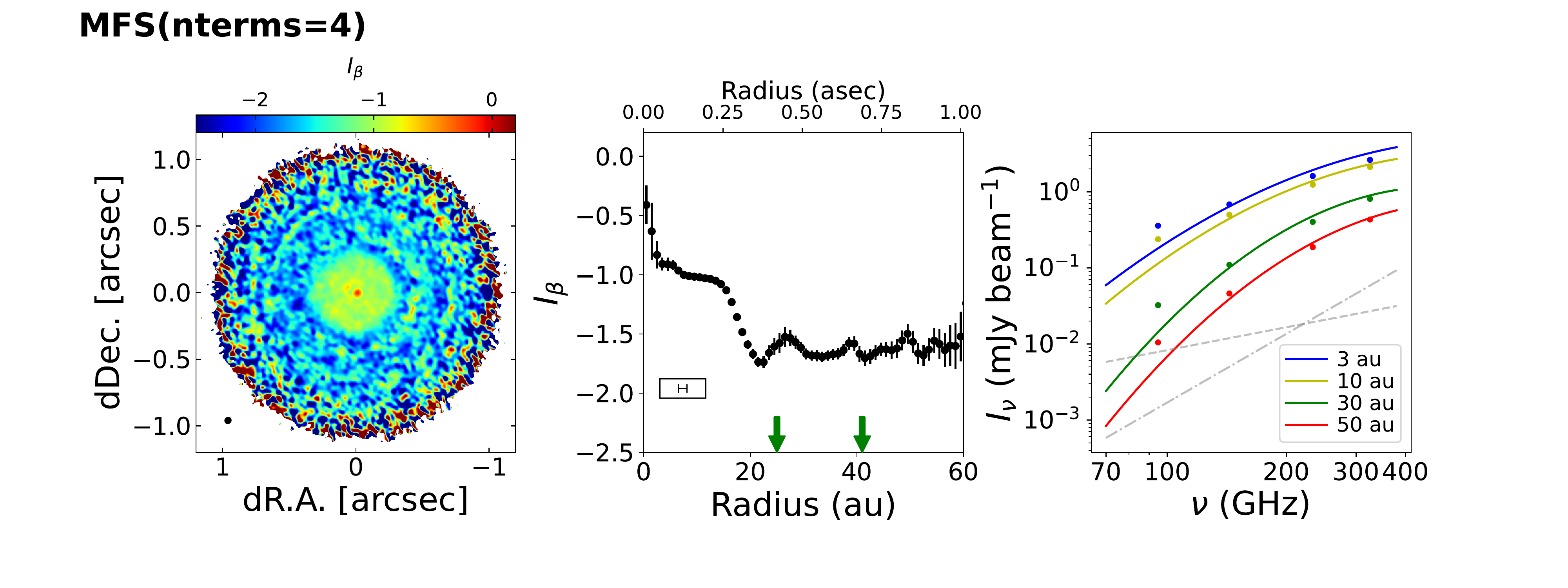}
\caption{Distribution of the spectral curvature $I_\beta$ and the inferred submillimeter spectrum for the cases of {\it nterms}=3 and 4 for the upper and lower panels, respectively. (Left) Map of $I_\beta$ reconstructed using the MFS method. The black circle at the bottom-left corner denotes the beam size. (Middle) Deprojected and azimuthally averaged profile of $I_\beta$. The error bars represent the standard error of the azimuthal averaging. The bar in the box at the bottom-left corner shows the geometric mean of the beam size. The green arrows in all the panels indicate the positions of the 25 and 41~au gaps. (Right) Submillimeter spectrum inferred from the $I_\alpha$ and $I_\beta$ profiles reconstructed with MFS. The azimuthally averaged spectrum at 10, 30, and 50~au are shown in blue, orange, and green, respectively. The black dots indicate the flux densities measured in each band map. The dashed and dash-dotted lines in gray show $\nu^2$ and $\nu^3$ dependence, respectively, as a reference.} \label{fig:MFS_beta}
\end{center}
\end{figure*}

If we adopt {\it nterms}=3 and 4, we can obtain a map showing the spectral curvature $I_\beta$ (Eq.~\ref{eq:mfs2}), as shown in Figure\ \ref{fig:MFS_beta}.
The maps of the spectral curvatures $I_\beta$ and their deprojected profiles clearly show that $I_\beta$ varies with radius for both cases.
Non-zero $I_\beta$ implies a frequency dependence in the spectral slope within the observed bands.
There is a difference in the value of $I_\beta$ between {\it nterms}=3 and 4 cases.
For {\it nterms}=3, $I_\beta$ is $\sim0$ near the disk center and gradually decreases to $-1.0$ toward the outer disk with a slight variation in the intensity gaps.
This indicates that, in almost all regions of the disk, the spectral index decreases as the frequency increases.
The positive $I_\beta$ at the innermost part of the disk implies the opposite trend of the spectral index, which is consistent with the existence of free-free emission at the stellar position suggested in previous studies \citep{bib:pascucci2012,bib:macias2021}.
On the other hand, for {\it nterms}=4, $I_\beta$ is from $-1.0$ to $-0.5$ near the disk center and drastically decreases to $\sim-1.7$ at $\sim$20~au.
The relatively large error bars of $I_\beta$ for the {\it nterms}=4 case is probably because a larger number of Taylor coefficients must be determined for higher orders of the Taylor expansion.
The submillimeter spectrum inferred from the $I_\alpha$ and $I_\beta$ profiles using Eq.~\ref{eq:mfs1} is also shown in Figure\ \ref{fig:MFS_beta}.
When comparing the flux densities of each band, it is clear that the combination of $I_\alpha$ and $I_\beta$ for the {\it nterms}=3 case reproduces the observed submillimeter spectrum better than the {\it nterms}=4 case.
This difference is because of the number of Taylor terms to describe the submillimeter spectrum.
In MFS, the submillimeter spectrum is described by Eq.~\ref{eq:mfs1} using three imaging parameters $I_0$, $I_\alpha$, and $I_\beta$, and they are calculated using the first three Taylor coefficients, $I_0$, $I_1$, and $I_2$.
With MFS {\it nterms}=3, the spectrum is described with the first three Taylor coefficients, and thus it is preferable because the parameters of the spectrum can be determined uniquely.
For {\it nterms}=4, on the other hand, we obtain four Taylor coefficients from the MFS imaging.
However, the final Taylor term is not employed for the spectrum calculation, though it is non-zero. 
This likely causes a difference between the calculated spectrum and the observed flux density for the case of {\it nterms}=4.

\begin{figure*}
\begin{center}
\plotone{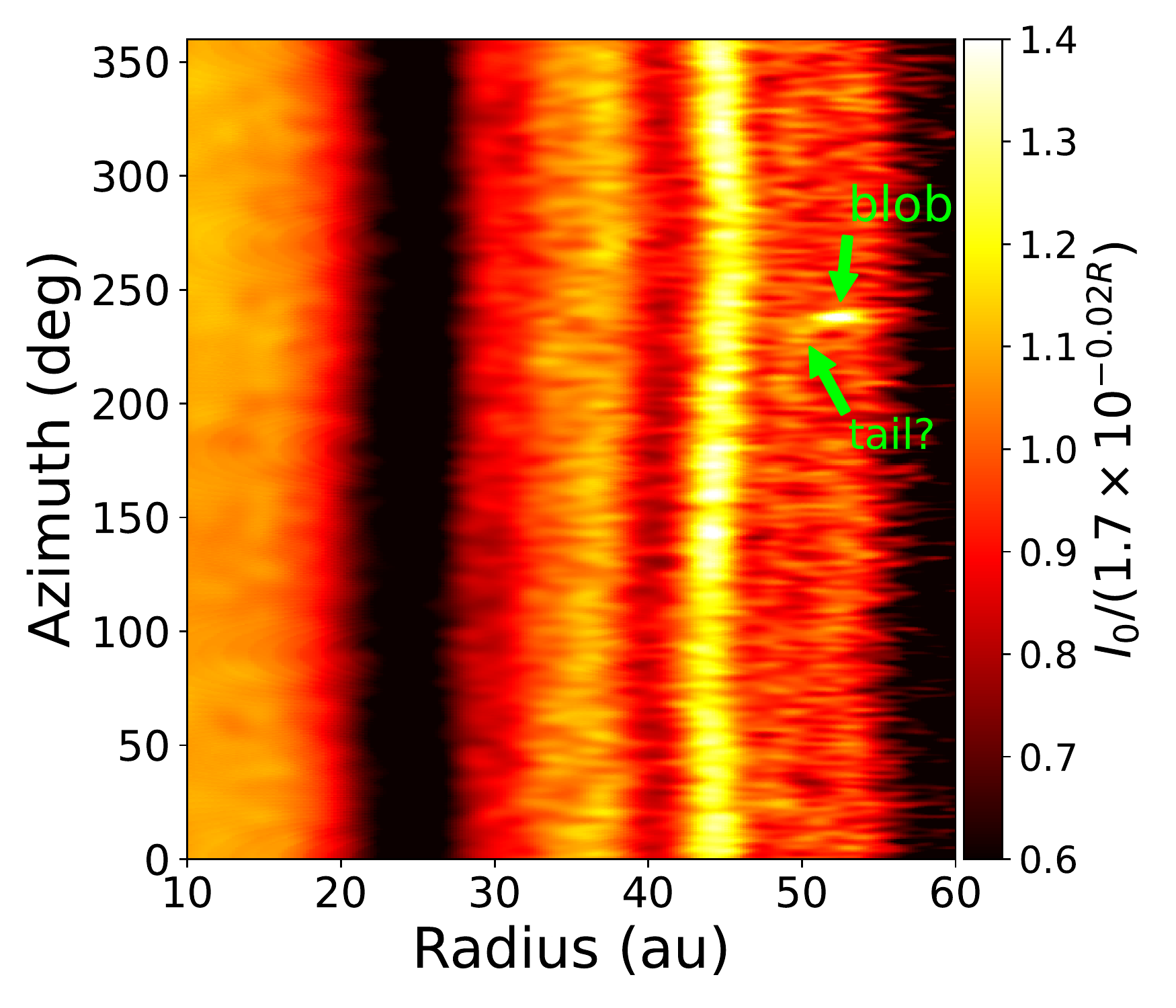}
\caption{Intensity map of MFS {\it nterms}=3 deprojected onto a polar coordinates. The intensity scale is divided by $1.7\times10^{-0.02R}$ mJy~beam$^{-1}$ to clearly view the features embedded in the background emission of the protoplanetary disk (see \S\ref{sec:model}). The millimeter blob emission reported by \citet{bib:tsukagoshi2019b} and a candidate trailing tail found in this study are labeled.}\label{fig:tail}
\end{center}
\end{figure*}

The point source sensitivity of our millimeter continuum image reconstructed with MFS is improved by $\sim$30 \% from the deepest one so far for TW Hya at high-resolution ($<$50~mas) by \citet{bib:tsukagoshi2019b}.
The high-resolution and high-sensitivity continuum map reconstructed with MFS provides an opportunity to search for substructures associated with the millimeter blob located at 52~au, as found by \citet{bib:tsukagoshi2019b}.
Figure \ref{fig:tail} shows the intensity map of MFS {\it nterms}=3 deprojected into a map in polar coordinates, whose intensity scale is normalized by an exponential function (see \S4) to more easily identify substructures.
We tentatively find an emission feature that could be a trailing tail that is emerged from the millimeter blob \citep{bib:nayakshin2020}.
However, it is also possible that the emission feature is an artifact caused by the residual of the sidelobe pattern. 
Another emission feature is that the emission ring at 45~au contains azimuthal wiggles, while dust wiggles are not present at the 33~au ring and 25~au gap.
This indicates that the 45~au ring has non-zero eccentricity or that the center of the ring orbit is slightly different to that for the inner rings/gaps.
Alternatively, the inner and outer disks might not be coplanar, or there might be an azimuthal variation of the ring scale height \citep{bib:doi2021}.
These emission features will be confirmed and discussed through future observations and more detailed analysis.

\epsscale{0.8}

\section{Validation of the Imaging Methods Using Intensity Models}\label{sec:model}
Our results indicate that for the MFS imaging, a higher-order Taylor expansion is required to reconstruct a reliable $I_\alpha$ map from datasets with wide frequency coverage at millimeter/submillimeter wavelengths.
The higher orders of Taylor expansion, however, require a significant SNR of the data.
On the other hand, although the resolution of the image is poorer than that of MFS, the image-oriented method provides an $I_\alpha$ map without using the Taylor series approximation for the frequency dependence.

In this section, we investigate the behavior of the MFS method using an intensity model to validate the reconstructed spectral index maps.
The intensity model is motivated by the intensity distribution of the TW~Hya disk.
The combined intensity and $I_\alpha$ maps were created using the same procedure as for the observed data.
We compared them to determine which is the more reliable procedure to make the $I_\alpha$ map from datasets with a wide frequency coverage.
Note that, for simplicity, we ignored the frequency dependence of the spectral slope, i.e., the spectral curvature.

\begin{figure}[htb]
\begin{center}
    \plotone{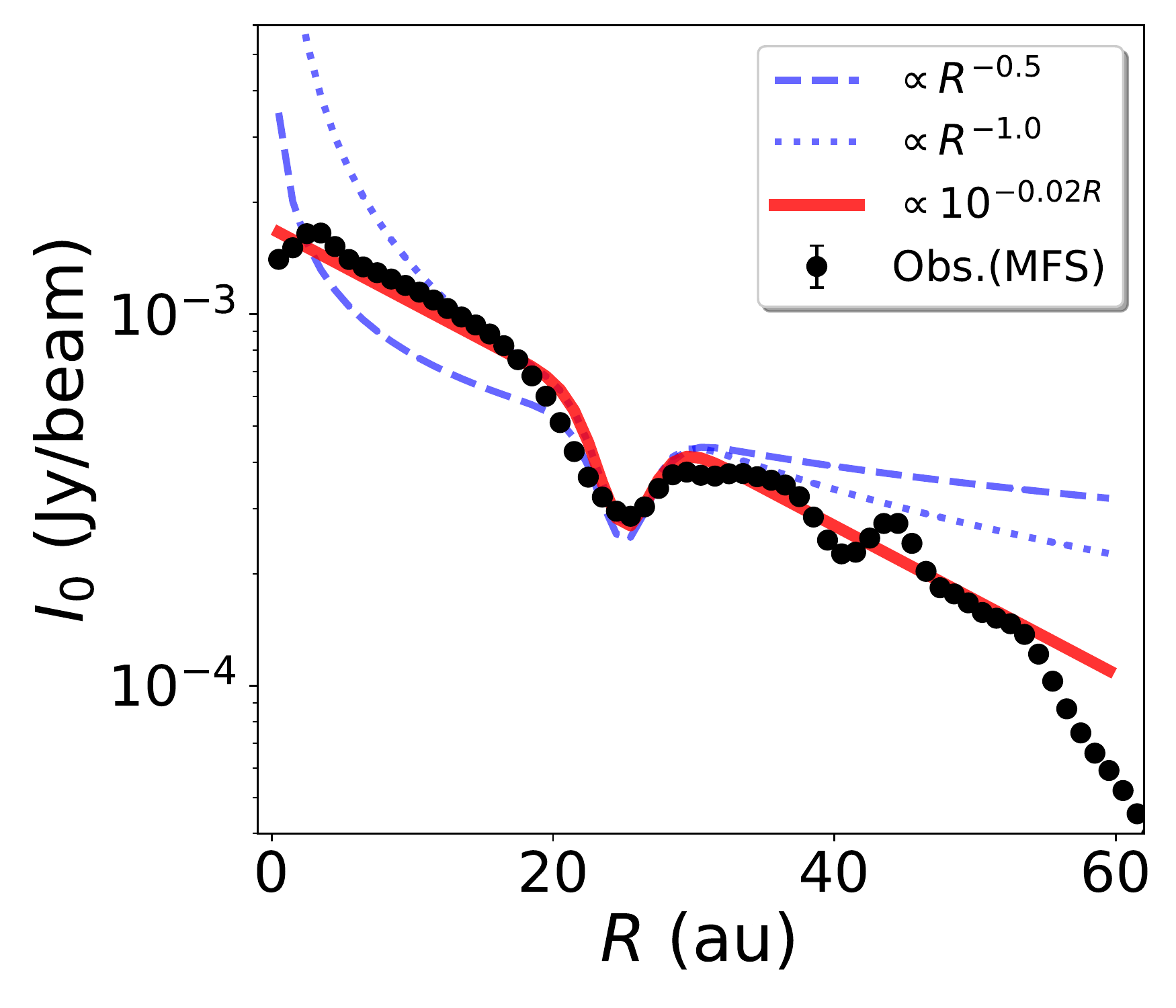}
    \caption{Comparison of the model intensity profile (red) to that of the observations (black dots). The radial profile of the combined intensity map reconstructed by MFS with {\it nterms=3} is adopted for the observed profile. The blue lines show a standard power-law form with $R^{-p}$ dependence.}\label{fig:model_profile}
\end{center}
\end{figure}

The intensity model was assumed to be an exponential function, as described by $I=1.7\times10^{-0.02R}$ mJy~beam$^{-1}$ at a representative frequency, i.e., the central frequency ($\sim$221~GHz).
The intensity profile was truncated at 1 and 60~au for the inner and outer radii, respectively.
We also added an intensity gap at 25~au to the model profile to more closely resemble that of the TW~Hya disk.
The gap is modelled using a Gaussian function with a FWHM of 5~au and a fractional depth of 0.5.
Figure \ref{fig:model_profile} shows the comparison of the adopted intensity distribution and the observed intensity.
The intensity profile of the model is more similar to the observed profile than the standard power-law dependence ($I\propto R^{-p}$).
For the radial dependence of the spectral index $\alpha(R)$, we assumed two cases.
One is a constant over the disk with a value of 2.5.
The other is a linear dependence with disk radius, in which $\alpha$=2 at 10~au and 3 at 50~au are assumed.

We assumed three cases for the radial dependence of the spectral index $I_\alpha(R)$.
The first one is a constant over the disk with a value of 2.5.
The second one is a linear dependence with disk radius, in which $I_\alpha$=2 at 10~au and 3.0 at 50~au are assumed.
Finally, we adopted the linear dependence assumed above with an enhancement at the 25~au gap.
The enhancement has a Gaussian form with the same width as the intensity gap (5~au in FWHM).
The peak value of $I_\alpha$ enhancement is set to be 3.

Under these assumptions, model images were created at the same frequency sampling as the observed datasets.
The model images were converted to visibilities and resampled to match each of the observations.
The visibilities were resampled using the Python code {\it vis\_sample}\footnote{https://github.com/AstroChem/vis\_sample} \citep{bib:loomis2017}.
Then, the model visibilities were imaged with the same parameters as for the observed datasets using the {\it tclean} task of CASA.

\begin{figure*}[htb]
\begin{center}
    \epsscale{0.9}
    \plotone{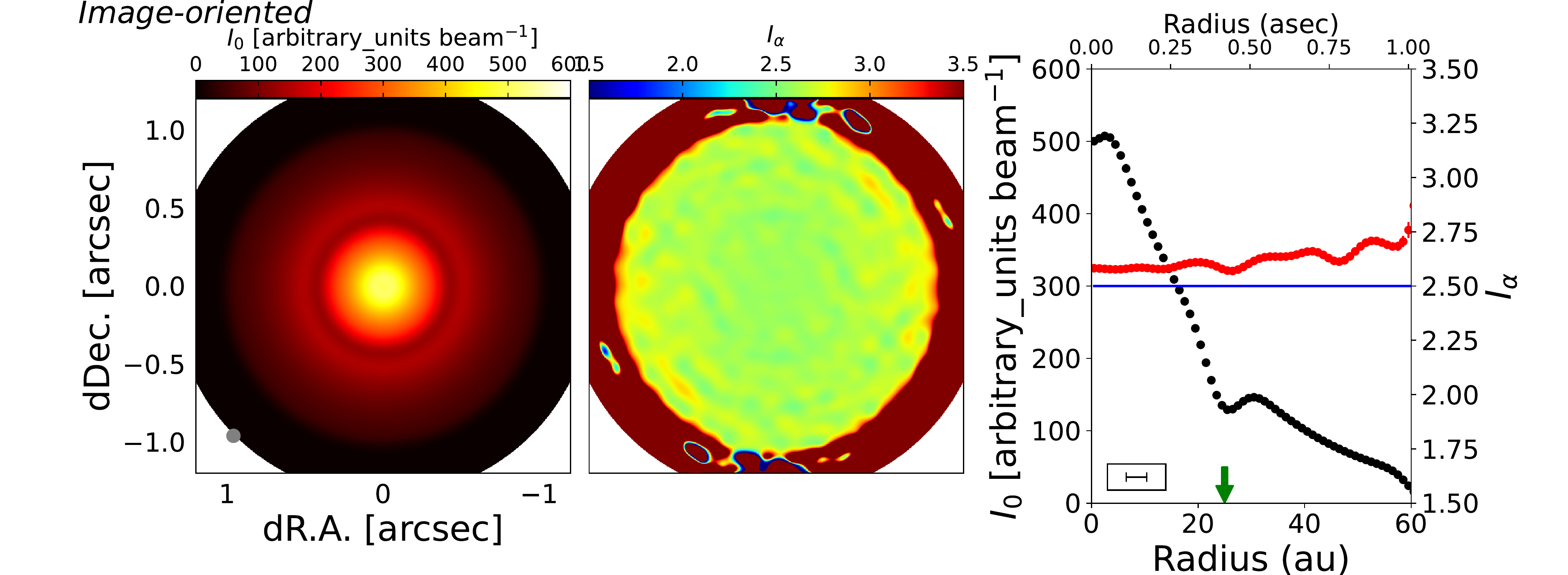}
    \plotone{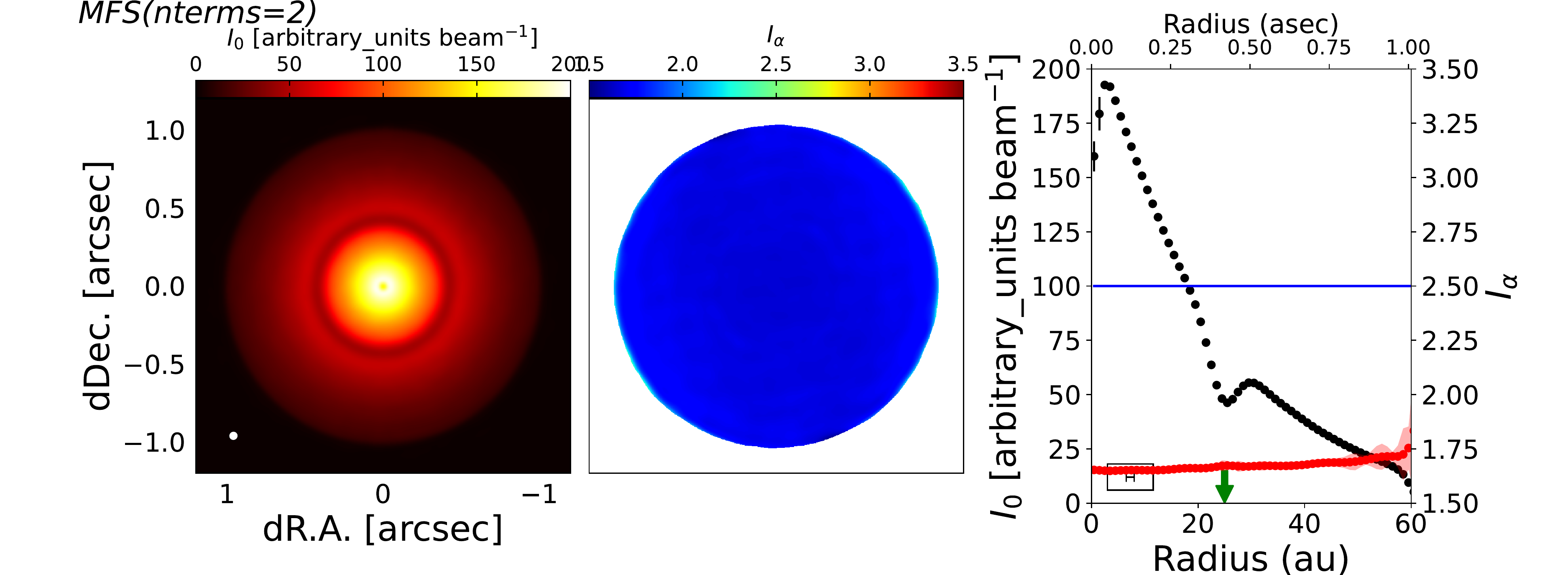}
    \plotone{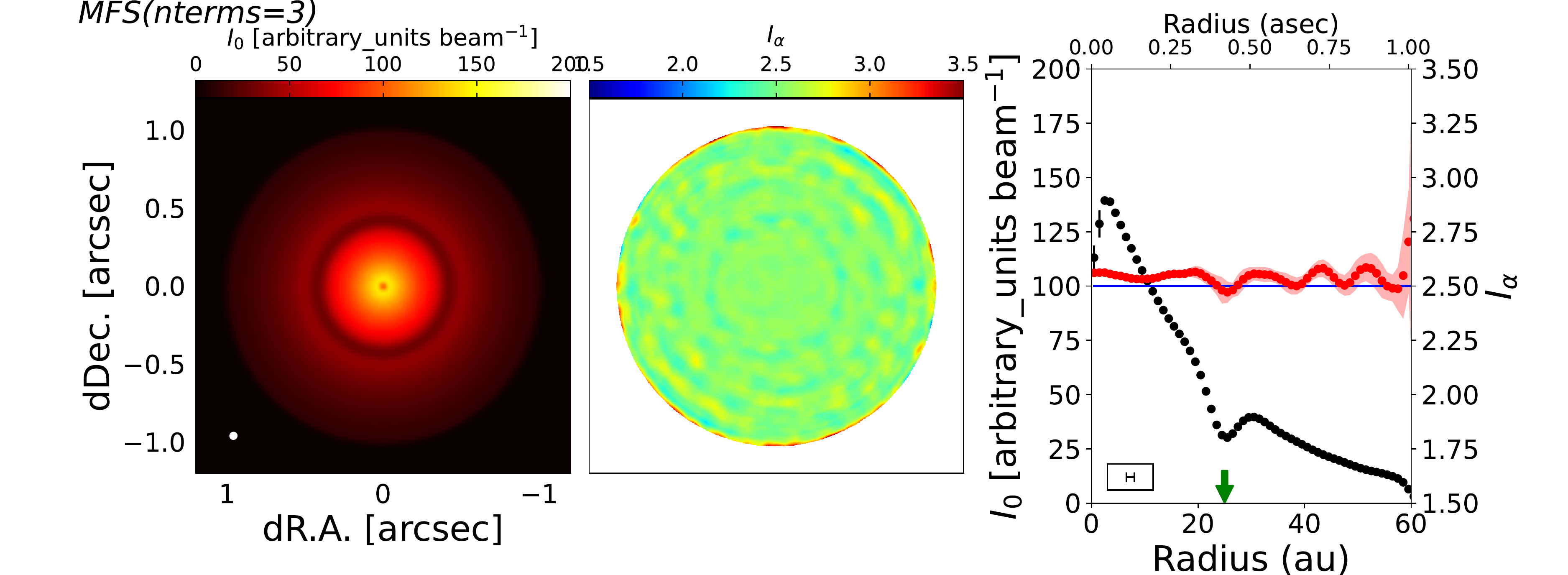}
    \plotone{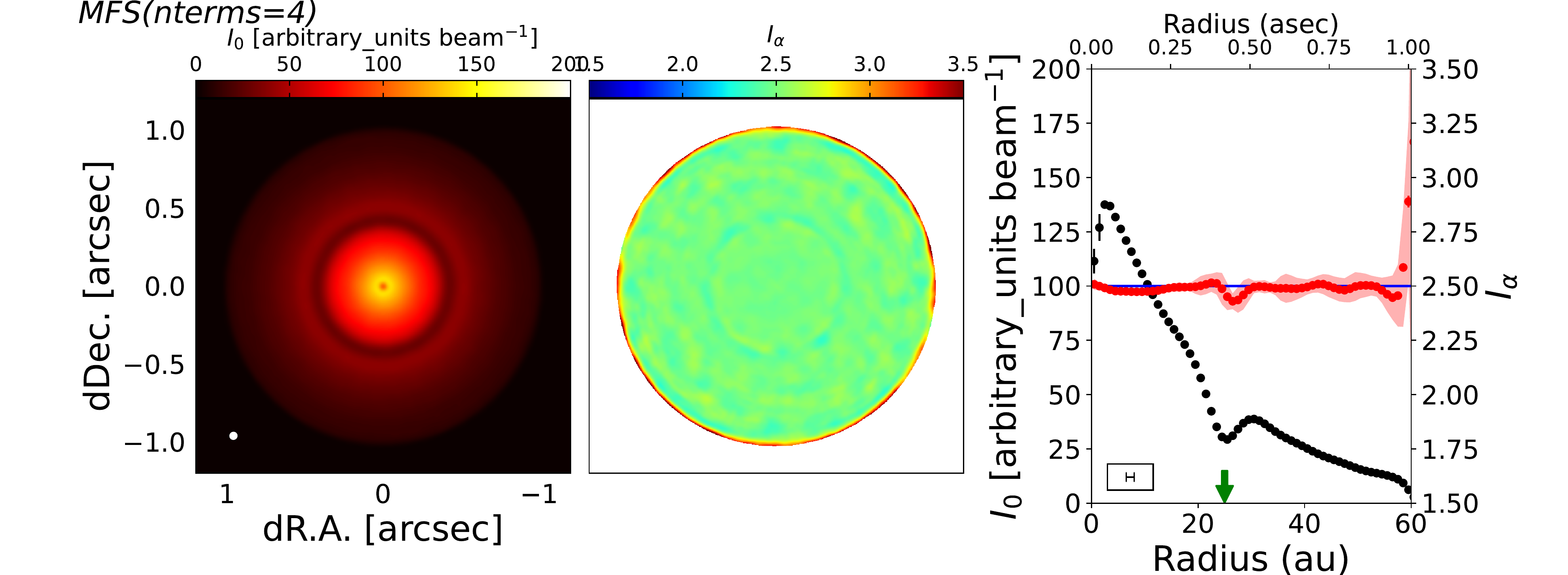}
    \caption{Comparison of the reconstructed images and radial distributions from the simulated visibilities using the image-oriented analysis (top) and the MFS method (from 2nd to bottom rows). The figure description is the same as that in Figure\ \ref{fig:spindex}. The intensity scale is in arbitrary units. The results for the constant $I_\alpha$ model are shown. The green arrow indicates the gap position in the model profile. The blue line represents the model $I_\alpha$ profile.}\label{fig:sim_result1}
\end{center}
\end{figure*}

\begin{figure*}[htb]
\begin{center}
    \epsscale{0.9}
    \plotone{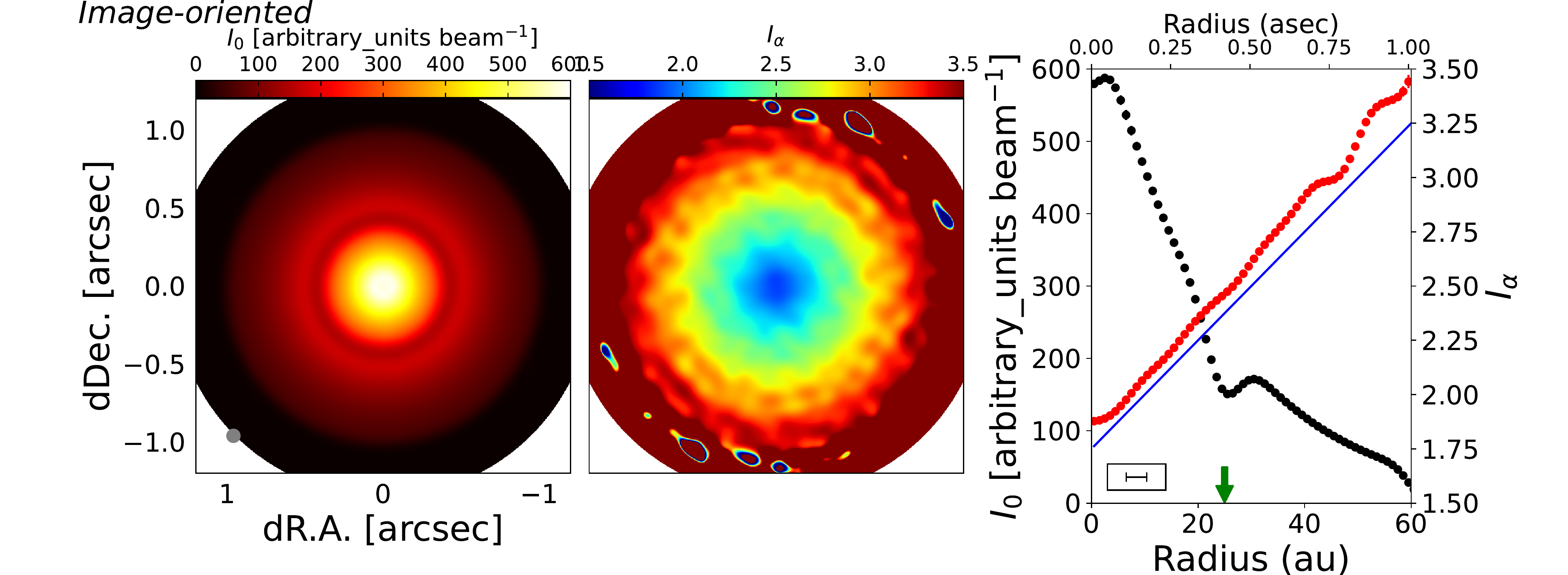}  
    \plotone{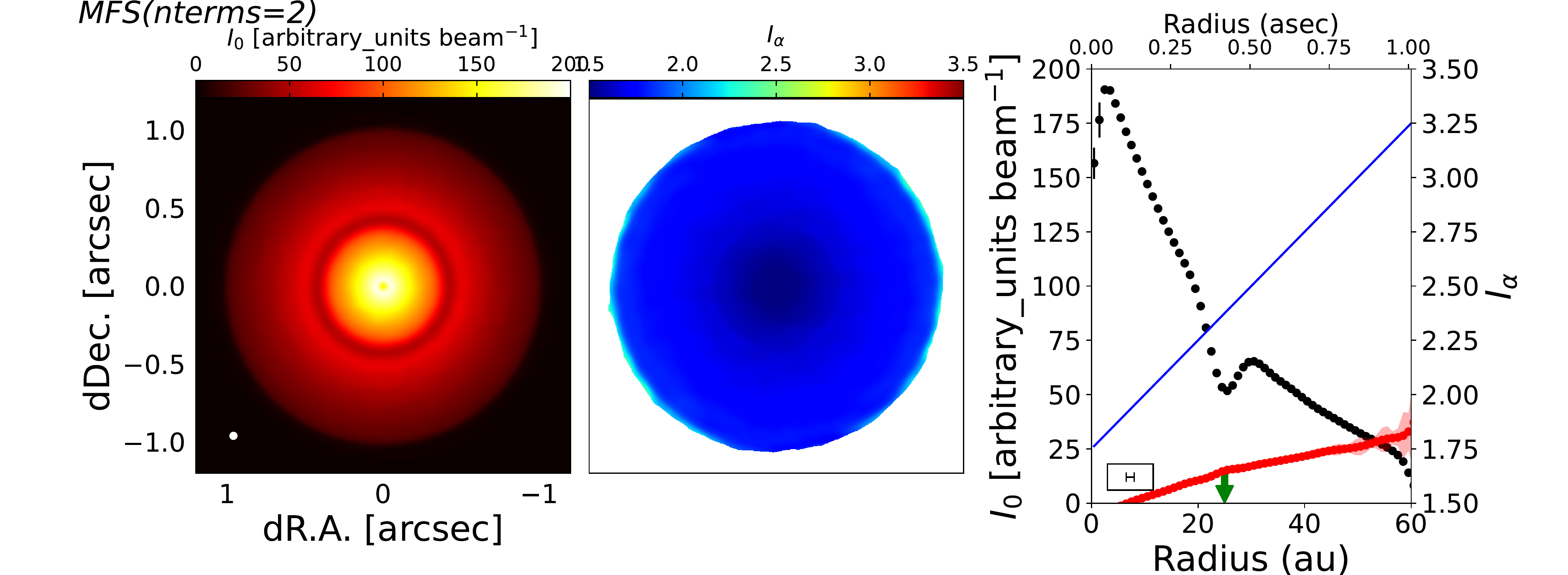}
    \plotone{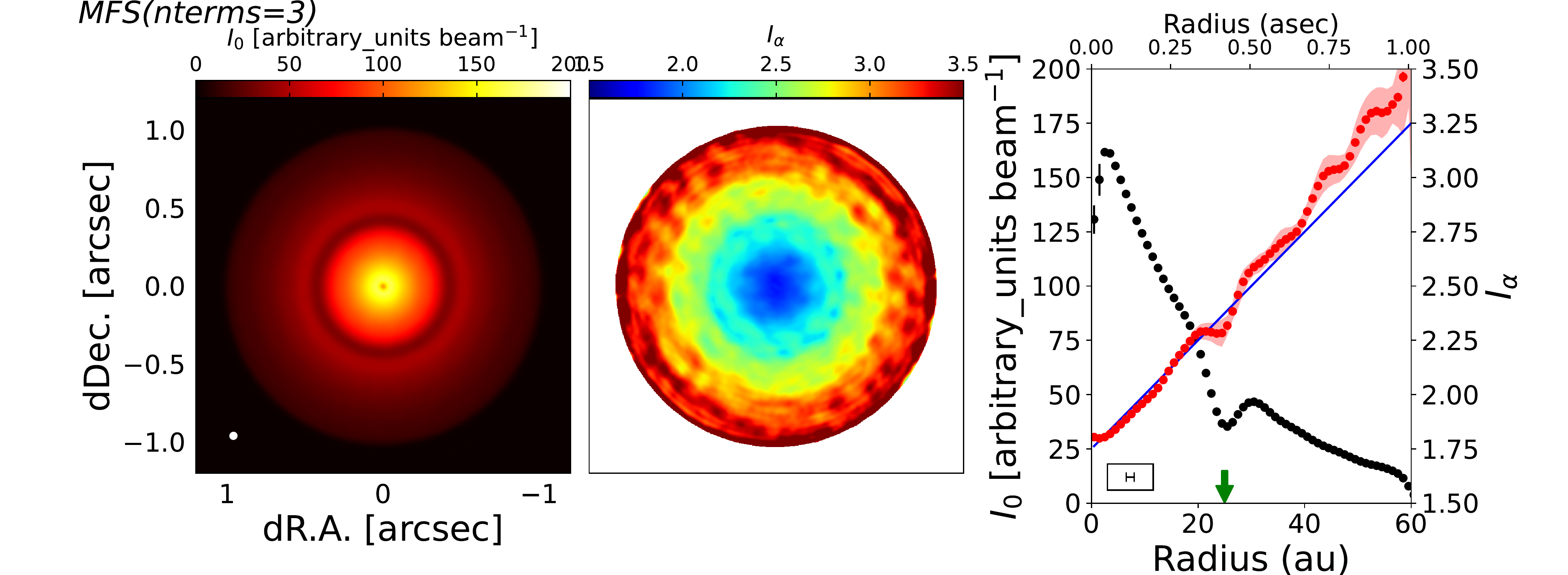}
    \plotone{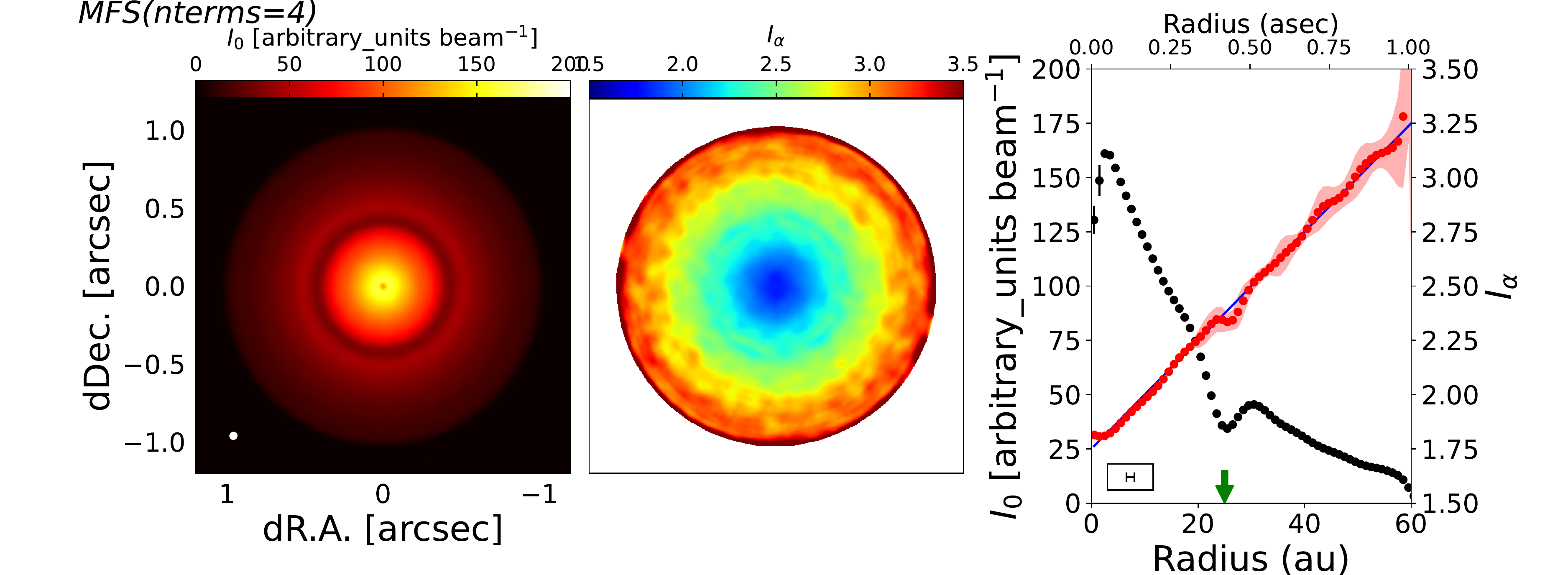}
    \caption{Same as Figure\ \ref{fig:sim_result1}, but for the gradient $I_\alpha$ model.}\label{fig:sim_result2}
\end{center}
\end{figure*}

\begin{figure*}[htb]
\begin{center}
    \epsscale{0.9}
    \plotone{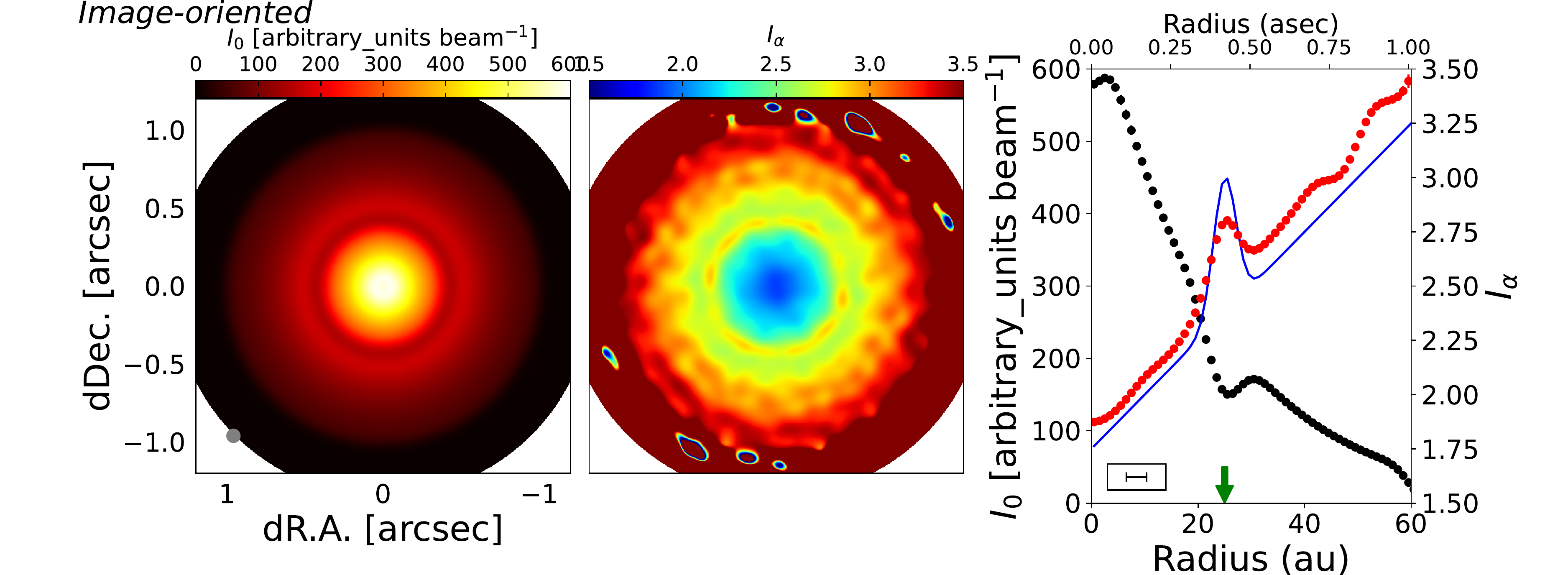}  
    \plotone{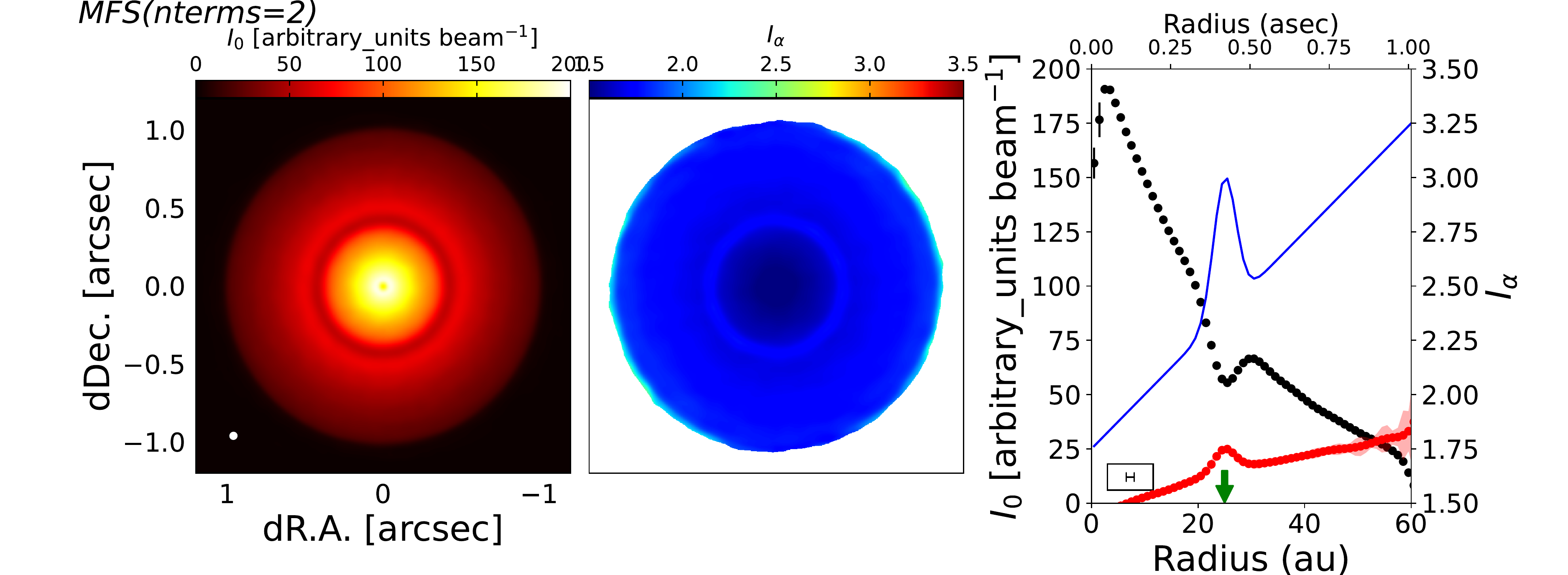}
    \plotone{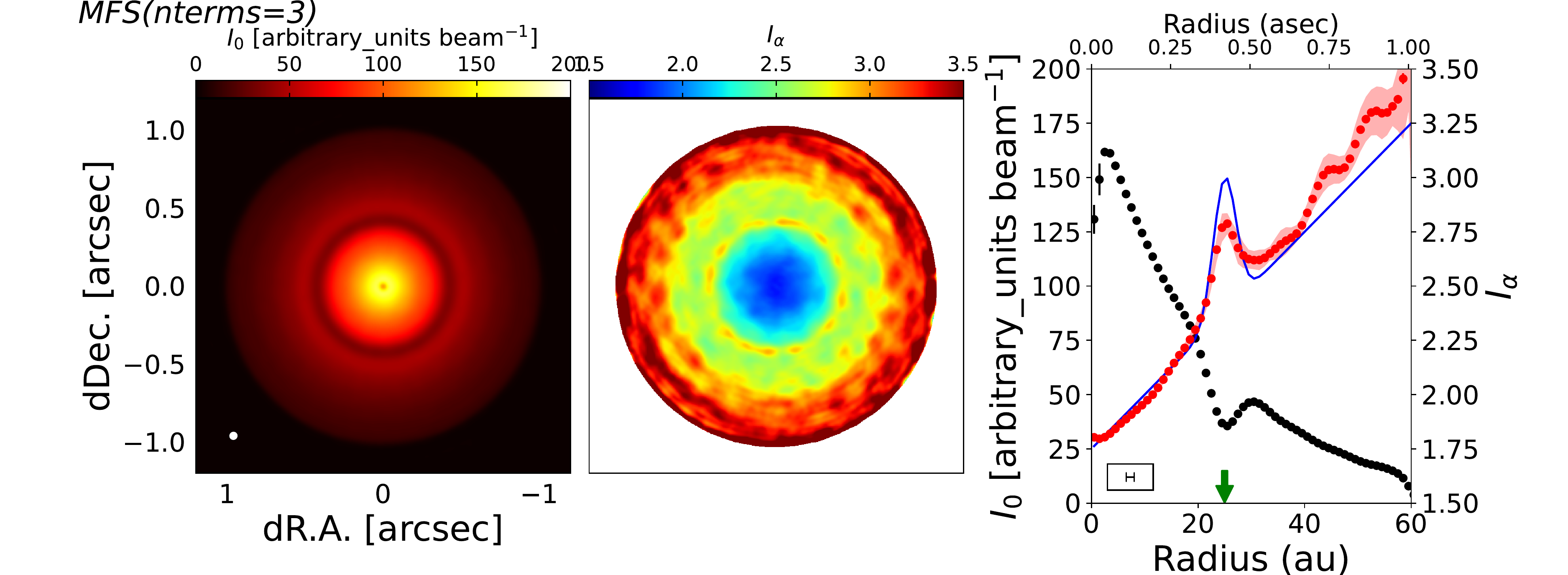}
    \plotone{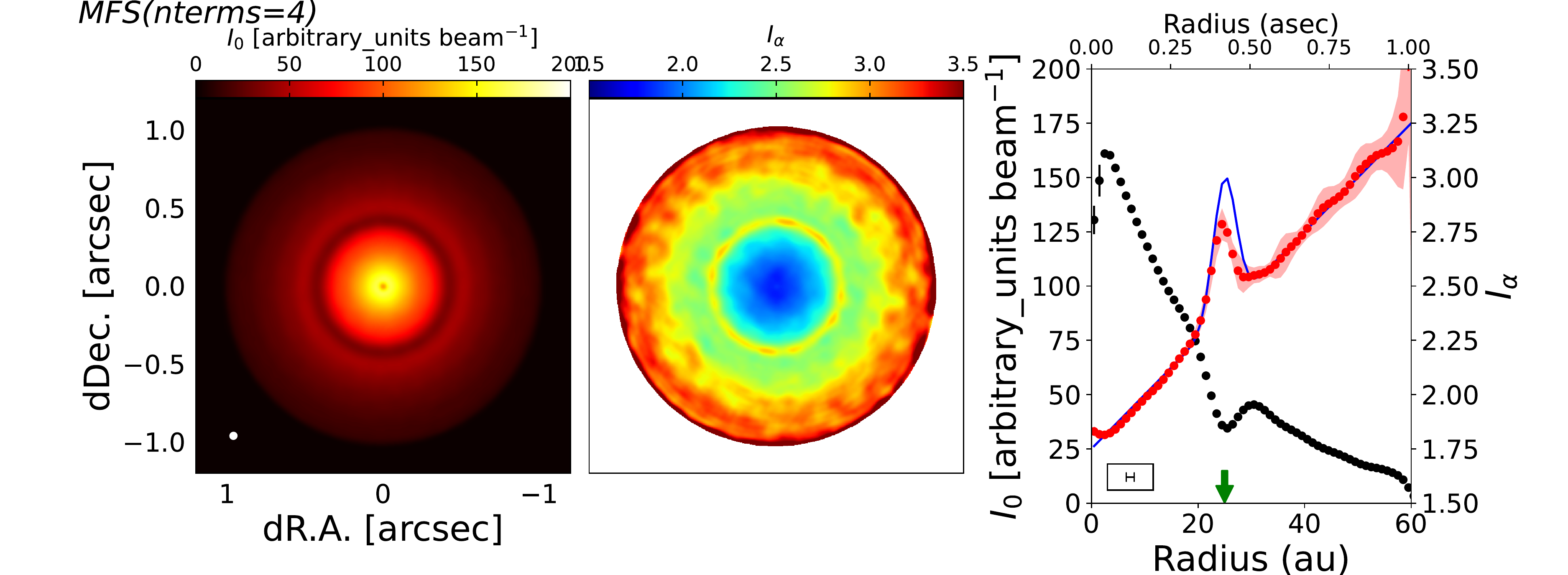}
    \caption{Same as Figure\ \ref{fig:sim_result1}, but for the model of the gradient $I_\alpha$ with an enhancement at 25~au.}\label{fig:sim_result3}
\end{center}
\end{figure*}

Figures\ \ref{fig:sim_result1}, \ref{fig:sim_result2}, and \ref{fig:sim_result3} compare the simulated images reconstructed from the model visibilities.
The reconstructed images of the intensity, spectral index, and their radial profiles are shown from left to right, respectively.
The results of the image-oriented method and of MFS with {\it nterms}=2, 3, and 4 are displayed from top to bottom.
We summarize the results of the imaging tests below.

\begin{itemize}

\item{
As mentioned in \S 2, the first order of the Taylor expansion ({\it nterms}=2) cannot reproduce the spectrum between Bands 4 and 7. 
The simulated $I_\alpha$ is $\sim$20\% lower than the input value for both the $I_\alpha$ models. 
The intensity of the combined map is also affected.
By adopting the first order of the Taylor expansion for the MFS imaging, the intensity at the central frequency tends to be overestimated by 20-40\% (see Figure \ref{fig:model_sed}). 
}

\item{
Following the above mentioned case, the MFS imaging that adopts {\it nterms}=3 and 4 reasonably reproduces the $I_\alpha$ profile, not only for the constant $I_\alpha$ case but also for the structured $I_\alpha$ cases. 
The difference between the mean $I_\alpha$ and the input value was typically less than 5\%. 
In addition, there appear to be artificial ripples over the disk with a spatial scale of $\sim$10~au, seen particularly in the case of {\it nterms}=3. 
Because the peak positions of the ripples vary if we adopt different multiscale parameters in CLEAN, the ripples could be caused by the combination of scale parameters.
}

\item{ 
Despite the difference in the $I_\alpha$ maps, the combined intensity is not significantly affected when {\it nterms}=3 or 4 is adopted.
In all the $I_\alpha$ model cases, the difference between the peak intensities of {\it nterms}=3 and 4 is less than 5\%, indicating that both profiles describe the radial distribution of the disk emission well.
}

\item{
In both the imaging methods, the existence of the intensity gap does not significantly affect the $I_\alpha$ profile.
If we adopt {\it nterms}=3 and 4, only a $<$2\% variation around the 25~au gap is found when the linear dependence of $I_\alpha$ is the case.
If there is an enhancement of $I_\alpha$ at the gap, the peak value of $I_\alpha$ is underestimated.
However, the difference is as small as $\sim$10\%.
Beam smearing could also be a reason for the decrease in $I_\alpha$ in the image-oriented method.
}

\item{
The image-oriented method is a good method to reproduce a reliable $I_\alpha$ map, although image resolution is sacrificed. 
The radial dependence of $I_\alpha$ agrees reasonably well with the input one, and the noise level is significantly lower than that of higher-order MFS images. 
One concern is that, in all the $I_\alpha$ model cases, the values of the simulated $I_\alpha$ profiles are slightly larger than the model ($\lesssim6$\%). 
This could be caused by the imaging of each band's intensity without using MFS because the bandwidth ratio of each type of data is not negligible. 
Alternatively, how deep we take clean components to make each band's image may also affect the spectral index map.
}

\end{itemize}

Based on these results, we conclude that MFS with the second order of the Taylor expansion ({\it nterms}=3) is a reasonable method to create a high-resolution combined intensity map.
This is because {\it nterms=2} cannot reproduce the flux density correctly because of the wide frequency coverage, and {\it nterms}=4 or higher causes difficulty in reconstructing the spectral curvature.
Although {\it nterms}=4 can provide a spectral slope comparable to or better than {\it nterms}=3, the number of Taylor coefficients is larger than the number of parameters required for describing a submillimeter spectrum ($I_0$, $I_\alpha$, and $I_\beta$), as shown in \S\ref{sec:results}.
The artifact of the $I_\alpha$ profile associated with the intensity gaps is negligible.
However, if $I_\alpha$ is enhanced at the gap, the peak value is underestimated by $\sim$10\%.
Although the resolution is lower than that of the MFS images, the image-oriented method provides a more robust $I_\alpha$ map.
The uncertainty owing to the selection of the imaging method is expected to be $\lesssim10$\% if the spectral curvature is negligible.
Thus, we conclude that the imaging method is reliable in checking both the $I_\alpha$ images reconstructed from MFS ({\it nterms=3}) and the image-oriented method.

Finally, we checked how the uncertainty in the absolute flux density calibration affects the $I_\alpha$ profile using the same procedure as the mock observation for the intensity model.
We employed an intensity model whose spectral index increases linearly with radius with an enhancement at the gap.
To observe the effect of the absolute flux calibration uncertainty on the reconstructed spectral slope, we ran mock observations for four cases in which the flux density of the model profile was modified by $\pm$10\% at Band~7 and $\pm$5\% at Band~3 and keeping the original $I_\alpha$ profile.
The flux densities at Bands~4 and 6 were unchanged.

\begin{figure*}[htb]
\begin{center}
    \epsscale{1.0}
    \plotone{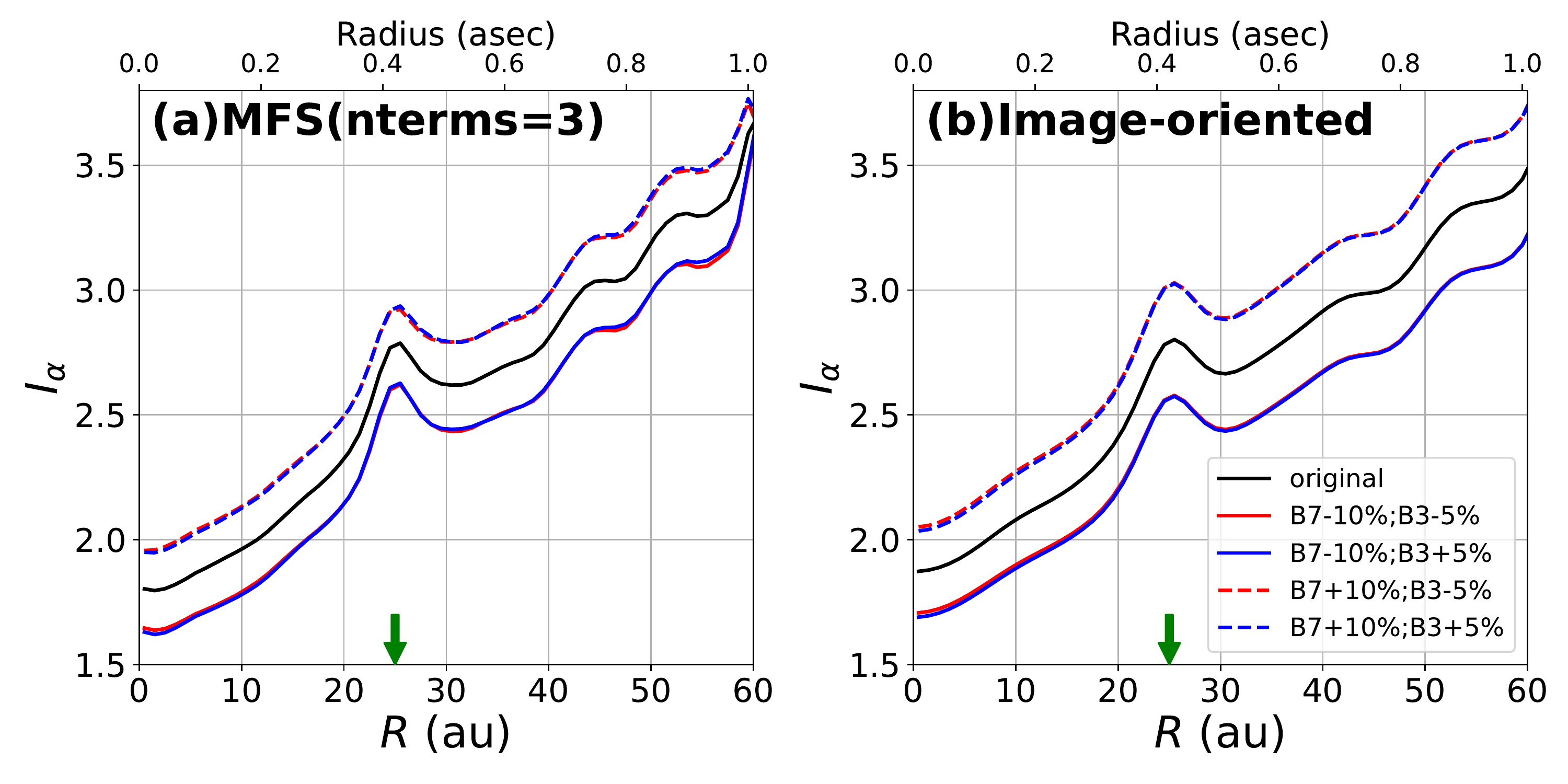}  
    \caption{Reconstructed $I_\alpha$ profiles simulated by adding the modulation of the flux density scale. The cases of $+10$\% and $-$10\% flux density modulation at Band~7 are shown in solid and dashed lines, respectively, and the cases of $+$5\% and $-$5\% modulation at Band~3 are shown in blue and red, respectively. The original profile without flux density modulation is shown in black. The green arrow indicates the gap position in the model profile. The results for MFS({\it nterms}=3)(a) and the image-oriented method (b) are shown.}\label{fig:flux_calibration}
\end{center}
\end{figure*}

Figure\ \ref{fig:flux_calibration} shows the results of the reconstructed radial profile of $I_\alpha$ using MFS({\it nterms}=3) and the image-oriented method.
It is clear that the uncertainty of the flux calibration does not affect the shape of the $I_\alpha$ profile, but does affect the value of $I_\alpha$.
Moreover, the value of $I_\alpha$ is more dependent on the uncertainty of the Band~7 flux calibration than that of Band~3.
The differences to the original profile are typically 7\% for the MFS method and 8\% for the image-oriented method.
Note that this is a conservative technique to determine the uncertainty due to the absolute flux calibration because we combine multiple measurement sets for each band.
Thus, the uncertainty of the absolute flux calibration should be lower than those reported by ALMA (10\% for Band~7 and 5\% for Band~3).

\section{Discussion} \label{sec:discussion}

\subsection{Comparison with the spectral index distribution of \citet{bib:macias2021}}
\begin{figure*}[htb]
\epsscale{1.0}
\begin{center}
\plotone{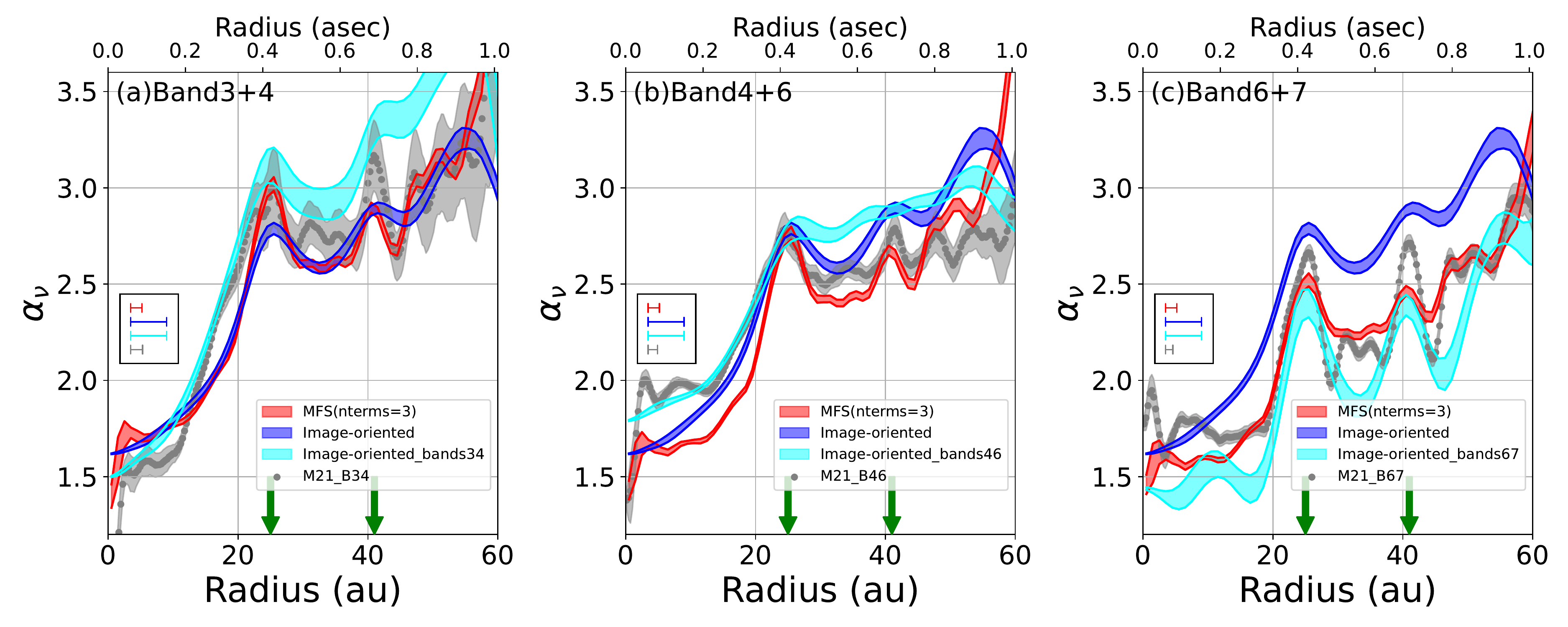}
\caption{Radial profile of the spectral slope $\alpha_\nu$ for the case of MFS ({\it nterms}=3) is shown in red. The profiles at the frequencies between Bands~3 and 4 (Band3+4;121~GHz), Bands~4 and 6 (Band4+6;190~GHz), and Bands~6 and 7 (Band6+7;290~GHz) are shown in (a)--(c), respectively.
The $\alpha_\nu$ profile determined by the image-oriented method using data from all bands is shown in blue, while those derived from the band-to-band fitting are shown in cyan.
The shaded area indicates the standard error.
The green arrows in the right panels indicate the positions of the 25 and 41~au gaps.
For reference, the radial profile of the spectral index measured in \citet{bib:macias2021} are shown by gray dots.
The bars in the box represent the beam sizes of the profiles.
}\label{fig:compare_spindex}
\end{center}
\end{figure*}

With $I_0$, $I_\alpha$, and $I_\beta$ derived with MFS {\it nterms}=3 (see Figures \ref{fig:spindex} and \ref{fig:MFS_beta}), we can describe the submillimeter spectrum using Eq.~\ref{eq:mfs1} and measure the spectral slope $\alpha_\nu$ at a specific frequency.
Figure\ \ref{fig:compare_spindex} shows the derived $\alpha_\nu$ at frequencies of 121, 190, and 290~GHz, which correspond to the central frequencies between ALMA Band~3 and 4 (Band3+4), 4 and 6 (Band4+6), and 6 and 7 (Band6+7), respectively.
The overall trend is that $\alpha_\nu$ decreases as the frequency increases.
This trend is more prominent at $\gtrsim$20~au; $\alpha_\nu$ decreases to a value of $\sim$0.5 from 121 to 290~GHz and $\sim$0.2 near 10~au, respectively.
The enhancements of $\alpha_\nu$ at the intensity gaps (25 and 43~au) appear in all $\alpha_\nu$ cases, and their excess compared to the surroundings decreases as the frequency increases.
As the spectral index determined by the image-oriented method using data from all bands is independent of the frequency, it seems to agree with the profile for the Band3+4 case, but cannot describe the profile for the Band6+7 case.
If we make $\alpha_\nu$ profiles of the image-oriented method using band-to-band fitting, the same trend in frequency as the MFS profiles is found although they have larger uncertainty.

Recently, \citet{bib:macias2021} presented the distribution of $\alpha_\nu$ for the TW~Hya disk at a resolution of 50~mas.
The difference from our study is that they measured $\alpha_\nu$ between Band~3--4, 4--6, and 6--7 separately by using MFS {\it nterms}=2, while our study focuses on determining $I_\alpha$ and $I_\beta$ through the MFS {\it nterms}=3 imaging.
Figure\ \ref{fig:compare_spindex} also compares our results of $\alpha_\nu$ with those derived by \citet{bib:macias2021}.
Our $\alpha_\nu$ profiles can reproduce the frequency dependence of \citet{bib:macias2021}.
The radial variation of the profiles is also almost consistent.
However, there is still a discrepancy in the excess in $\alpha_\nu$ at the intensity gaps; the results of \citet{bib:macias2021} show that the excess of $\alpha_\nu$ at the gaps is largest at Bands 6 and 7, whereas our result shows the opposite trend.
This is probably because our $\alpha_\nu$ measurement by combining data over four bands improves the signal-to-noise ratio of the profile.

\subsection{Implication of the dust size distribution}
In this subsection, we deduce the optical depth $\tau_0$ at the central frequency $\nu_0$, the power-law index of the dust opacity $\beta$, and the temperature of the dust disk $T_\mathrm{d}$ using the submillimeter spectrum derived from our MFS imaging.
If we neglect the scattering of dust, the intensity of the dust emission $I_\nu$ is expressed as
\begin{equation}
I_\nu = B(T_\mathrm{d}) \left[ 1-\exp \left\{-\tau_0 \left(\frac{\nu}{\nu_0} \right)^\beta \right\} \right],
\end{equation}
where $B(T_\mathrm{d})$ is the Planck function.
There are three unknown variables, $\tau_0$, $\beta$, and $T_\mathrm{d}$ in this equation.
On the other hand, the observed submillimeter intensity, $I_\nu \mathrm{(obs)}$, can be expressed using three parameters determined by our MFS imaging $I_{\nu_0}$, $I_\alpha$, and $I_\beta$ as
\begin{equation}
I_\nu \mathrm{(obs)}= I_{\nu_0} \left( \frac{\nu}{\nu_0} \right) ^{I_\alpha + I_\beta \log \left( \frac{\nu}{\nu_0} \right)}.
\end{equation}
This implies that we can solve the unknown three variables from the submillimeter spectrum.

To address this problem, we calculated the minimization of $\Delta I_\nu \equiv I_\nu - I_\nu\mathrm{(obs)}$ by varying $\tau_0$, $\beta$, and $T_\mathrm{d}$.
We used {\it curve\_fit} in the {\it scipy} optimization module to minimize $\Delta I_\nu$.
To prevent divergence, the solution is searched with the minimum and maximum bounds of 0.001--10, -3--3, and 10--300 for $\tau_0$, $\beta$, and $T_\mathrm{d}$, respectively.
The standard errors in the radial profiles of the observed parameters were used to determine the weight of the minimization.

\begin{figure*}[htb]
\begin{center}
\epsscale{1.0}
    \plotone{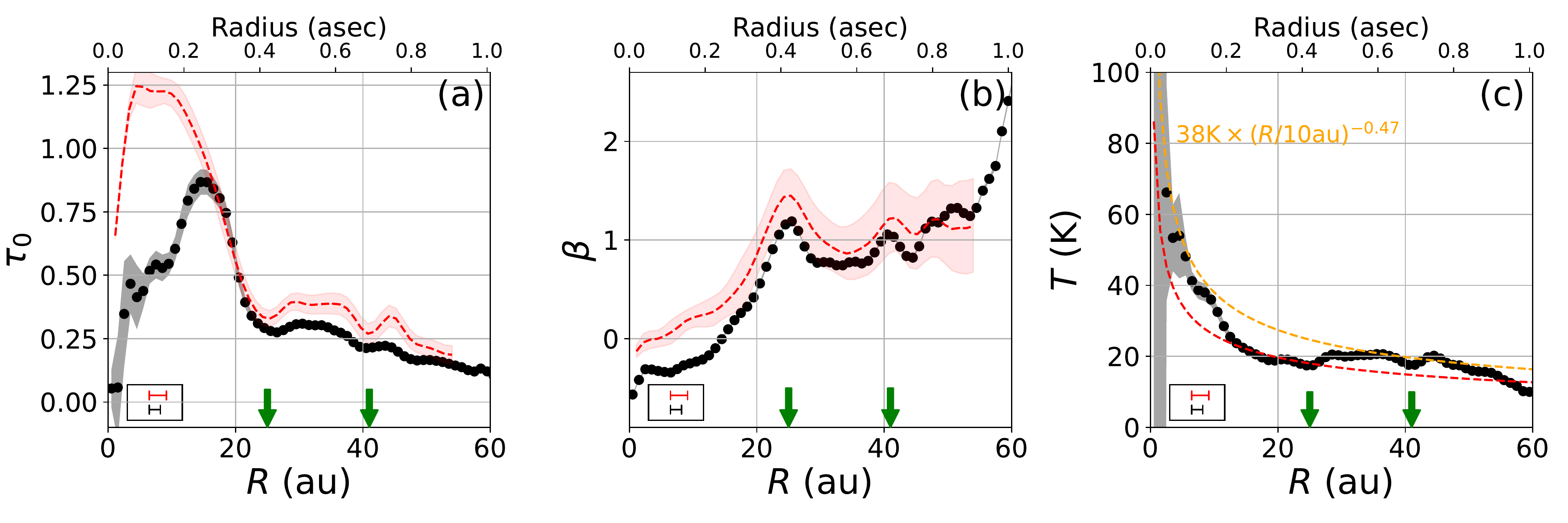}
    \caption{Radial profiles of the optical depth at 221~GHz $\tau_0$ (a), power-law index of the opacity coefficient $\beta$ (b), and temperature of the dust disk $T_\mathrm{d}$ (c). The profiles are calculated based on the maps reconstructed by MFS {\it nterms=3}. The shaded region corresponds to minimization error. As a reference in (a) and (b), $\tau$ and $\beta$ at 190~GHz derived by \citet{bib:tsukagoshi2016} are shown by the dashed red line with the shaded region showing the standard error ($T_\mathrm{10}=$26~K and $q=-0.4$ case). In (c), the disk midplane temperature profiles assumed in \citet{bib:tsukagoshi2016} and derived by \citet{bib:zhang2017} are indicated by the red and yellow dashed lines, respectively. The green arrows in all the panels indicate the positions of the 25 and 41~au gaps. The mean beam sizes of this study and \citet{bib:tsukagoshi2016} are shown by the bars in black and red, respectively, at the bottom-left corner of each panel.}\label{fig:tau-beta}    
\end{center}
\end{figure*}

The derived profiles of $\tau_0$, $\beta$, and $T_\mathrm{d}$ are shown in Figure\ \ref{fig:tau-beta}.
The disk is entirely optically thin at 221~GHz, although only marginally so at $\sim$15~au.
The shape of the $\tau_0$ profile is consistent with that obtained in \citet{bib:tsukagoshi2016} at $>$15~au, but it deviates at the inner radii mainly due to the difference in disk temperature profiles adopted.
As shown in Figure\ \ref{fig:tau-beta}(c), our direct measurement of the dust temperature $T_\mathrm{d}$ agrees well with the estimate obtained by a modeling approach for the gas disk \citep{bib:zhang2017}.

The radial dependence of $\beta$ is similar to that derived in \citet{bib:tsukagoshi2016}.
The value is slightly smaller ($\sim$0.1--0.3) than that derived in \citet{bib:tsukagoshi2016}, probably because of the difference of the frequency range over which $\beta$ was determined.
The enhancements of $\beta$ associated with the 25 and 41~au gaps are also seen.
This result still supports the conclusion of \citet{bib:tsukagoshi2016}, which is that this can be explained by a deficit of millimeter-sized grains within the gap.
In the inner region of the disk ($R\lesssim$15~au), where the effect of the optical depth cannot be ignored, $\beta$ is less than 0; this indicates that the emission is blackbody-like, or that the scattering of millimeter radiation is effective \citep{bib:liu2019,bib:ueda2020}.
The scattering should be responsible for an approximately one order of magnitude difference between our estimate of the optical depth and that derived by \citet{bib:macias2021}.

According to theoretical predictions of the dust opacity \citep[e.g.,][]{bib:birnstiel2018}, $\beta \sim 1$ implies that the power-law index of the dust size distribution $q$ is $\sim$ 3.5 and that the maximum dust particle size is above 1~mm.
Beyond the 25~au gap, where the emission is optically thin, $\beta$ varies up to $\sim$1.5 at 60~au, meaning that the maximum dust size could be a few millimeters. 

This conclusion is supported by the detailed modeling of the dust size distribution for sets of high-resolution ALMA data \citep{bib:macias2021}.
Note that, in our study, the disk parameters are determined from optically thinner frequency bands (Bands~3, 4, 6, and 7).
By adding optically thick continuum data at higher frequency bands (Band~9 or 10), the disk parameters, particularly the dust temperature profile, can be determined more robustly \citep{bib:kim2019}.

\section{Summary}\label{sec:summary}
To obtain a higher-sensitivity intensity map at millimeter wavelengths and to revisit the dust size distribution of the protoplanetary disk around TW~Hya, we created high-resolution maps of the intensity and the spectral index by combining sets of ALMA data at Bands~3, 4, 6, and 7.
In addition to using the existing ALMA archive data, we have newly conducted high-resolution observations at Bands~4 and 6, a part of which has already been published in \citet{bib:tsukagoshi2019b}.
Two methods are employed to reconstruct the combined intensity and the spectral index maps; a traditional image-oriented method and multi-scale and multi-frequency synthesis (multi-scale MFS).
The impact of the choice of the methods was also investigated using an intensity model motivated by TW~Hya.
The results of this paper are summarized as follows:

\begin{itemize}

\item We show the spectral index map reconstructed with both imaging methods. A reasonable method to reconstruct the spectral index map is MFS with the second order of the Taylor expansion for the frequency ({\it nterms}=3). With a smaller order of the Taylor expansion ({\it nterms}=2), the number of Taylor coefficients is too small to reproduce the frequency dependence from Bands 3 to 7. Meanwhile, the higher-order ({\it nterms}=4) MFS imaging requires a larger number of Taylor coefficients and a higher signal-to-noise ratio. Although the resolution is almost twice as poor, the image-oriented method provides a consistent spectral index map with MFS ({\it nterms}=3) imaging.

\item The spectral index reconstructed with MFS {\it nterms}=3 agrees well with that derived in previous studies \citep{bib:tsukagoshi2016,bib:huang2018,bib:macias2021}. The index decreases toward the disk center and shows enhancements in the intensity gaps. The spectral index of the image-oriented method showed similar structures. Our MFS {\it nterms}=3 imaging shows that the submillimeter spectrum of TW~Hya has spectral curvature, indicating that the spectral index depends on the frequency.

\item We investigated how the substructures of intensity distribution affect the reconstructed spectral index map using an intensity model and noise-free mock observations. We validated that the first order of the Taylor expansion is insufficient to reproduce the frequency dependence from Bands 3 to 7, and the higher-order of Taylor expansion of MFS ({\it nterms}=3 and 4) is necessary. We found that the higher-order MFS method can provide a high-resolution spectral index distribution with an uncertainty of $<10$~\% and the presence of the intensity gap does not significantly influence the reconstruction of the spectral index distribution. Although the resolution is lower than that of the MFS images, the image-oriented method also provides a robust distribution of the spectral index if there is no frequency dependence in the spectral index.

\item We formulated the submillimeter spectrum of the TW~Hya disk as a function of the disk radius by using the images reconstructed with MFS {\it nterms}=3. With the spectrum, the optical depth $\tau_0$, power-law index of the opacity coefficient $\beta$, and temperature of the dust disk $T_\mathrm{d}$ were derived under the assumption that scattering is negligible. The derived $\tau_0$ and $\beta$ agree well with those derived in our previous work \citep{bib:tsukagoshi2016}. The enhancement of $\beta$ at the intensity gaps was also confirmed, supporting a deficit of millimeter-sized grains within the gap.

\item By combining all the visibilities from Bands~3 to 7, we made the highest sensitivity continuum map at millimeter wavelengths to date. The point source sensitivity of our map was improved by 30\% from the previous highest sensitivity continuum map of \citet{bib:tsukagoshi2019b}. The previously reported substructures in the dust emission were confirmed by our maps. The tentative detection of a new emission feature associated with the millimeter blob has also been reported, but it should be confirmed by future observations and detailed analysis.

\end{itemize}

\acknowledgments
We would like to thank the referee for improving our manuscript. We are also grateful to Enrique Macias for sharing the spectral index profiles of the TW~Hya disk.
This paper makes use of the following ALMA data: ADS/JAO.ALMA\#2013.1.00114.S, 2013.1.00387.S, 2015.A.00005.S, 2015.1.00308.S, 2015.1.00686.S, 2015.1.00845.S, 2016.1.00229.S, 2016.1.00311.S, 2016.1.00440.S, 2016.1.00464.S, 2016.1.00629.S, 2016.1.00842.S, 2016.1.01495.S, 2017.1.00520.S, and 2018.1.01218.S.
ALMA is a partnership of ESO (representing its member states), NSF (USA) and NINS (Japan), together with NRC (Canada), MOST and ASIAA (Taiwan), and KASI (Republic of Korea), in cooperation with the Republic of Chile.
The Joint ALMA Observatory is operated by ESO, AUI/NRAO, and NAOJ. 
A part of the data analysis was carried out on the common-use data analysis computer system at the Astronomy Data Center of NAOJ.
This work was supported by JSPS KAKENHI Grant Numbers 17K14244, 18H05438, and 20K04017.
CW acknowledges financial support from the University of Leeds, and the Science and Technology Facilities Council and UK Research and Innovation (grant numbers ST/T000287/1 and MR/T040726/1). TJM thanks the STFC for support via grants ST/P000312/1 and ST/T000198/1.
M.T. was supported by JSPS KAKENHI grant Nos. 18H05442,15H02063,and 22000005.
 
%

\vspace{5mm}
\facilities{ALMA},


\software{CASA \citep{bib:mcmullin2007},
          numpy \citep{bib:numpy2020}, 
          scipy \citep{bib:scipy2020}, 
          astropy \citep{bib:astropy2013}, 
          matplotlib \citep{bib:matplotlib2007}, 
          vis\_sample \citep{bib:loomis2017}}



\bibliography{refs.bib}



\end{document}